 \documentclass[reprint,NumberedRefs]{JASA}

\begin{document}

\title[JASA/Sample JASA Article]{Reduced order modelling using parameterized non-uniform boundary conditions in room acoustic simulations}
\author{Hermes Sampedro Llopis}
\email{hsllo@elektro.dtu.dk}
\affiliation{Acoustic Technology Group, Department of Electrical and Photonics Engineering, Technical University of Denmark,
Kongens Lyngby, Denmark}
\affiliation{Rambøll Denmark, Copenhagen, Denmark}

\author{Cheol-Ho Jeong}
\affiliation{Acoustic Technology Group, Department of Electrical and Photonics Engineering, Technical University of Denmark,
Kongens Lyngby, Denmark}

\author{Allan P. Engsig-Karup}
\affiliation{Scientific Computing Section, Department of Applied Mathematics and Computer Science,
Technical University of Denmark, Kongens Lyngby, Denmark}


\date{\today} 

\begin{abstract}
Quick simulations for iterative evaluations of multi-design variables and boundary conditions are essential to find the optimal acoustic conditions in building design. We propose to use the reduced basis method (RBM) for realistic room acoustic scenarios where the surfaces have inhomogeneous acoustic properties, which enables quick evaluations of changing absorption materials for different surfaces in room acoustic simulations. The RBM has shown its benefit to speed up room acoustic simulations by three orders of magnitude for uniform boundary conditions. This study investigates the RBM with two main focuses, 1) various source positions in diverse geometries, e.g., square, rectangular, L-shaped, and disproportionate room. 2) Inhomogeneous surface absorption in 2D and 3D by parameterizing numerous acoustic parameters of surfaces, e.g., the thickness of a porous material, cavity depth, switching between a frequency independent (e.g., hard surface) and frequency dependent boundary condition. Results of numerical experiments show speedups of more than two orders of magnitude compared to a high fidelity numerical solver in a 3D case where reverberation time varies within one just noticeable difference in all the frequency octave bands. 

\end{abstract}


\maketitle

\section{\label{sec:1} Introduction}
 Room acoustic simulations are typically used during the design stage of a building to find the optimal amount, choice and position of materials according to the use of the room. Poor acoustic conditions can produce a negative effect, e.g., a decrease in the work productivity \citep{officeproductivity1, officeproductivity2}, a decrease in the learning quality \citep{schools1,schools2}, and an increase in the stress level \citep{sundbyberg,officestress}. Performing room acoustic simulations by solving the acoustic wave equation using numerical methods, such as Finite Difference Time-Domain methods (FDTD)\citep{FDTD}, Finite Element Methods (FEM) \citep{FEM} or Spectral Element Methods (SEM)\citep{SEM} is accurate as all important wave phenomena can be accounted for. However, it is computationally expensive compared to geometric acoustics methods \citep{Geometr1}. As a result, active research seeks to find ways to speed up the numerical methods to be used in building design. Common strategies include using model order reduction (MOR) \citep{CRBM}, parallel computing on multi-core processing units \citep{parallel}, and more recently also machine learning techniques \cite{Nborrel2}. 
 
  A review of the state-of-the-art in MOR across disciplines can be found in the literature \cite{MORsota}. The reduced basis method (RBM) \citep{CRBM} is a method belonging to the class of MOR techniques. It exploits the parametric dependence in the solution of a partial differential equation (PDE) by combining different solutions given by the variation of a set of parameter values. Problems of large systems of PDEs are effectively reduced to a low dimensional subspace to achieve computational acceleration and transformed back to the original problem size \citep{CRBM2, NgocCRBM,RozzaHuynhRBM,DrohmannRBM, PrudRBM, QuarteroniRBM,HolmesTurb,RBMSuccessLimitations}, however, the speedup is dependent on the problem of interest. MOR have been recently presented in some acoustic applications \citep{Quirin1,Quirin2,rombem} as well as in many different fields, e.g., electromagnetics \citep{Ganeshele, Chenelec}, computational fluid dynamics \citep{AmsallemFluid, phdGiere}, heat transfer \citep{phdGrepl}, vibroacoustics \citep{SrinivasaVA, HetmaniukVA, HerrmannVA} and many others \citep{romart1,romart2,romart3,romart4,romart5}, demonstrating, in general, an efficient reduction in the computational burden. Using RBM in room acoustic simulations with parameterized boundary conditions during the design stage of an indoor space may allow exploring many acoustic conditions by solving the reduced system and exploiting the benefit of a reduced computational cost under the variation of several parameters, e.g., the thickness of a porous material, air gap distance, etc. Today, the application of MOR to room acoustic simulations with boundary parameterization is scarce despite the significant potential for acceleration. A recent study \citep{Miller2} presents a model order reduction strategy using a Krylov subspace algorithm in the time domain with a FEM solver where speedups of 11–36 were demonstrated. The study is performed for a simple domain where all the surfaces are assumed to be rigid boundaries, and only the floor is modelled with a surface impedance. Moreover, another study demonstrates the potential of RBM in room acoustic applications, however, only for simplified homogeneous boundary conditions \citep{RBMHermes}. Two to three orders of speedup factors were achieved in the online stage.
 
The study of RBM with realistic inhomogeneous boundary conditions, where the absorption materials are distributed inhomogeneously among the surfaces, is scarce in the literature. For example, classrooms typically have highly absorbing ceilings and scattering objects near the rear surface and window on one side wall. 
 In this study, we investigate how to effectively apply RBM in a realistic setting with inhomogeneous boundary conditions, how the RBM performance varies with the room geometry and source/receiver location, and what acceleration is expected in such conditions.
 
The novelty of this investigation is to construct and analyze the performance of a RBM strategy for realistic scenarios with two focuses, one being the RBM performance in various room shapes and source locations and two being the performance when including numerous parameters of acoustic materials distributed inhomogeneously. First, this study presents a conceptual proof-of-concept for a complex 2D case, as it simplifies the computational burden. Second, two 3D rooms are analyzed. This study is essential for future work, e.g., scaling the method to extend the RBM for large building projects.

In Sec. \ref{sec:2} the governing equations, the boundary conditions, the reduced basis method and the error measures are described together with the different domains and simulation parameters. Section \ref{sec:3} handles the numerical experiments and results, which are analyzed and discussed in Sec. \ref{sec:4}.

\section{\label{sec:2}Methods}
This section presents an overview of the methods and simulation conditions used in this study. A more detailed description of the full order model (FOM) solver used as a reference model and the ROM is deeply described in previous work \citep{RBMHermes}.

\subsection{Governing equations and boundary conditions}

We consider the acoustic wave equation in the Laplace domain
\begin{equation}\label{welap}
    s^2p- c^2 \Delta p = 0, 
\end{equation}

where $p(\boldsymbol{x},t)$ is the sound pressure, $\boldsymbol{x}\in \Omega$ the position in the domain $\Omega\subset 	\mathbb{R}^d$ with $d=\{2,3\}$, $t$ is the time in the interval $(0, T]$ s and $c$ is the speed of sound ($c = 343$ m/s).
Equation (\ref{welap}) is discretized using the SEM formulation, which is well known and an overview can be found  \citep{NODALDGFEM,SEM,SPHPAllan}. The final formulation written in the Laplace domain is given by
\begin{align}\label{eqdiscretized}
 \big(s^2\mathcal{M}+c^2\mathcal{S} + sc^2\frac{\rho}{Z_s}\mathcal{M}_{\Gamma}\big)\mathbf{p} =0,
\end{align}
where $\mathcal{M}$ refers to the mass matrix, $\mathcal{S}$ is the stiffness matrix,  $\rho$ is the density of the medium ($\rho=1.2$ kg/m$^3$) and $Z_s$ is the surface impedance. Note that the impedance boundaries are considered denoting the boundary domain as $\Gamma$.\\

The frequency dependent boundary conditions are implemented via the method of auxiliary differential equations (ADE) \citep{Cotte,Troian,SEM}. The surface impedance of a porous absorber is modelled using Miki's model \citep{Mikis} in conjunction with a transfer matrix method \citep{allardbook}, and mapped to a six pole rational function by using a vector fitting algorithm \citep{vecfitt} so that the surface admittance $Y_s=1/Z_s$ can be written as a rational function and expressed using partial fraction decomposition \citep{Troian}. Then, the system (\ref{eqdiscretized}) can be stated in the form of a linear system of equations and solved using a sparse direct solver as presented in \citep{SEM, RBMHermes} 
\begin{align}\label{eqsystem}
\mathbf{K}\mathbf{p} =0, \quad \mathbf{K}\in \mathbb{R}^{N \times N},
\end{align}
where, $\mathbf{K}$ refers to the operators shown in  (\ref{eqdiscretized}) and $N$ corresponds to the degree of freedom (DOF). Note that if the system is split into real and imaginary parts for implementation purposes, the size of the operator is $\mathbf{K}\in \mathbb{R}^{2N \times 2N}$ \citep{Bigoni, RBMHermes}. The system (\ref{eqdiscretized}) is initialized using a Gaussian pulse with a spatial distribution $\sigma_g$ that determines the frequencies to span, by adding the right hand side term $s\mathcal{M}p_0$, where $p_0$ is the initial sound pressure state in the time domain.

The solution in the Laplace domain is finally transformed to the time domain by means of the Weeks method \citep{Weeks,RBMHermes}.

\subsection{The reduced order model}
The purpose of using RBM is to substantially reduce the size of the problem while ensuring a certain level of accuracy. Specifically, the DOF are reduced, and the techniques succeed when RDOF $\ll$ DOF, RDOF being the corresponding degrees of freedom in the numerical scheme after applying RBM.
The RBM consists of two stages. First, one or more variables present in the partial differential equation or its discretized form (\ref{eqdiscretized}) are chosen as parameters, e.g., $Z_s$, spanning a discrete range of values. Then, in the first stage, referred to as the offline stage, the parameter space is explored to generate a problem-dependent basis by collecting FOM solutions for different parameter values within the range of interest. A Galerkin projection takes place to reduce the dimensionality of the problem by utilizing the generated basis. In the second stage, referred to as the online stage, the reduced problem is solved for a new parameter value that was not explored in the offline stage at a much lower computational cost. The offline stage is typically computationally costly as it requires multiple FOM solutions that capture relevant information on the parameters variations and allow generating of representative basis functions of that variation. The reduction of the size of the computational problem comes with the truncation of the basis. The solver is stated in the Laplace domain to ensure the stability of the ROM solution \citep{Bigoni}. The basis generation can be seen as a data-driven technique based on proper orthogonal decomposition (POD). It relies on a proper symplectic decomposition (PSD) with a symplectic Galerkin projection. Specifically, \textit{the cotangent-lift} method introduced in \citep{symplectic} is applied, which preserves the structure of the operators when the problem is split and solved into real and imaginary parts. The ROM solution is expressed as an expansion of the basis functions $\phi_i$ and coefficients $a_i$, which is represented as
\begin{align}\label{eqrom}
p_{rom} = \mathbf{\Phi}\mathbf{a}, 
\end{align}
where $\Phi_{ij} \equiv \phi_i(x_j)$. Inserting (\ref{eqrom}) into (\ref{eqsystem}) yields to a similar problem, where now the system is solved for the coefficients $\mathbf{a}$
\begin{align}\label{eqsystemrom}
\mathbf{K}_{rom}\mathbf{a} =0,
\end{align}
where $\mathbf{K}_{rom} = \boldsymbol{\Phi}_{cl}^T \mathbf{K}\boldsymbol{\Phi}_{cl}$ and $\boldsymbol{\Phi}_{cl}$ defines the symplectic basis constructed as
\begin{equation}
   \boldsymbol{\Phi}_{cl} = \begin{bmatrix} \boldsymbol{\Phi} & \boldsymbol{0} \\ \boldsymbol{0} & \boldsymbol{\Phi} \end{bmatrix}.
\end{equation}
Moreover, the reduced operator can be written also as
\begin{align}\label{eqsystemrom}
\mathbf{K}_{rom} = s^2\boldsymbol{\Phi}^T\mathcal{M}\boldsymbol{\Phi}+c^2\boldsymbol{\Phi}^T\mathcal{S}\boldsymbol{\Phi} + sc^2\frac{\rho}{Z_s}\boldsymbol{\Phi}^T\mathcal{M}_{\Gamma}\boldsymbol{\Phi},
\end{align}
where a new parameter value can be chosen during the online stage  (for example, $Z_s$). The generation of a basis comes by first collecting all the FOM solutions obtained during the offline stage in the snapshot matrix\citep{RBMHermes} $\boldsymbol{S}_N  \in \mathbb{R}^{N\times 2N_{s}}$, where $N$ is already described above, and $N_s$ is the number of evaluated complex frequencies determined by the Weeks method \citep{Bigoni}. $\boldsymbol{S}_N$ can be decomposed using the proper orthogonal decomposition (POD) technique based on a singular value decomposition (SVD) to get the corresponding basis functions, defined as  $\boldsymbol{\Phi} = [U_1,...,U_{N_{rb}}]\in \mathbb{C}^{N\times N_{rb}}$. The reduced basis is chosen through truncation of this basis relying on the rate of decay of the singular values.\\

The singular value decay shows the energy distribution among the basis, providing information about the reduction of the problem. It is defined as 
\begin{equation}
    E/E_0 = \frac{\textrm{diag}(\sigma_1,..,\sigma_N)}{\sum_{i=1}^N \sigma_i},
\end{equation}
where $\Sigma=\textrm{diag}(\sigma,..,\sigma_N)$ is obtained by SVD where $S=U\Sigma V^T$. The number of  basis can be chosen so that the projection error is smaller than a given tolerance $\epsilon_{POD}$

\begin{equation}\label{romNrbselec}
    I(N_{rb}) = \frac{\sum_{i=1}^{N_{rb}} \sigma^2_i}{\sum_{i=1}^N \sigma^2_i} \geq 1-\epsilon_{POD},
\end{equation}
where $N_{rb}$ denotes the number of basis functions, $I(N_{rb})$ represents the percentage of the energy of the collection of FOM solutions captured by the first. For a given $\epsilon_{POD}$, the faster the energy decays, the smaller number of basis needed, and thus, a better reduction of the problem is expected. The reduction comes with a truncation of the basis, which determine the size of (\ref{eqsystemrom}) as $N_{rb}$.\\

\subsection{Error measures}
In this study, two type of errors are considered to compare the ROM against the FOM. First, the relative  error using the root mean square (rms) pressure is introduced
\begin{equation}\label{eq:errel}
    \epsilon_{rel}=\frac{p_{rms_{ROM}}-p_{rms_{FOM}}}{p_{rms_{FOM}}}\times 100 \quad (\%).
\end{equation}
Second, the error in the frequency domain expressed in dBs is considered
\begin{equation}\label{eqerrorL}
    \Delta L (f)=20\log_{10}\Big|\frac{p_{FOM}(f)}{p_{ROM}(f)}\Big|,
\end{equation}
where, $p_{FOM}$ and $p_{ROM}$ are the sound pressure of the FOM and ROM respectively along the frequency spectrum.

The performance of the ROM is measured in terms of speedups ($sp$) defined as
\begin{equation}\label{eqspeedup}
    sp=\frac{CPU_{FOM}}{CPU_{ROM}},
\end{equation}
where $CPU_{FOM}$ and $CPU_{ROM}$ corresponds to the computational time of the FOM and ROM respectively.

\subsection{Test rooms and simulation conditions}
\begin{figure}
\fig{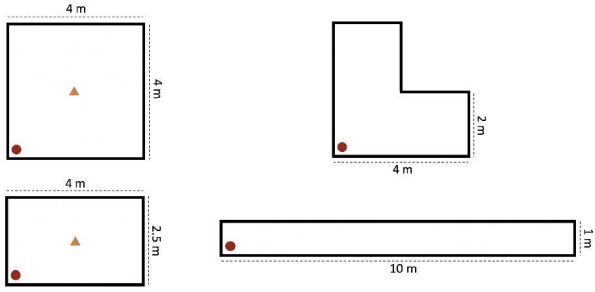}{8cm}{}
\caption{The geometries and source positions.}
\label{fig1geom}
\end{figure}

First, Section \ref{sec:influece}, deals with ROMs with several 2D geometries with different source locations illustrated in Figure \ref{fig1geom}. Four different geometries and two source positions are considered, one at the corner and one at the centre. First a $4\times 4$ m$^2$ square domain with the source placed at $(s_{x_1},s_{y_1})_{SQ}=(0.2, 0.2)$ m (SQ1), and $(s_{x_2},s_{y_2})_{SQ}=(2, 2)$ m (SQ2) is introduced (Figure \ref{fig1geom}a). Second, a  $4\times 2.5$ m$^2$ rectangular domain where  $(s_{x_1},s_{y_1})_{RC}=(0.2, 0.2)$ m (RC1), and $(s_{x_2},s_{y_2})_{SQ}=(2, 1.25)$ m (RC2) is considered (Figure \ref{fig1geom}b). Third, an L-shaped room where the long side is $4$ m and the short side is $2$ m is considered with only one source position at the corner  $(s_{x},s_{y})_{LS}=(0.2, 0.2)$ m (LS1) (Figure \ref{fig1geom}c). Finally a corridor shape of size $10\times 1$ m$^2$ is presented where  $(s_{x},s_{y})_{CO}=(0.2, 0.2)$ m (CO1) (Figure \ref{fig1geom}d).
The maximum element size is selected considering triangular high-order elements ($P=4$) and using $4$ points per wavelength ($PPW$) leading into an upper frequency $f_u=2.8$ kHz, which is approximately the upper cutoff frequency of the $2$ kHz octave band. The model is excited with a Gaussian pulse as initial condition with $\sigma_g=0.1$ m$^2$ \citep{Sakamoto}.
The ROMs for each room type and source position are built by sampling the surface impedance of all the sides at the following values $Z_s= [500, 5250, 10000]$ Nsm$^{-3}$. 

Second, section \ref{sec:param2d} deals with several 2D ROMs for an inhomogeneous distribution of the acoustic material, which in this study is named inhomogeneous boundary conditions. A 2D rectangular room ($4$ m $\times 2.5$ m) with inhomogeneous boundary conditions is considered with the same simulation parameters as before. The source is placed at  $(s_{x},s_{y})=(3, 1.2)$ m and the receiver is at  $(r_{x},r_{y})=(1, 1.2)$ m. Moreover, an additional ROM is constructed for this section with an upper frequency of $f_u=4$ kHz.

Third, for section \ref{sec:param3d}, two 3D models are considered. A $1$ m cube (CB) enclosure is presented, where three of the six surfaces are parameterized. Simulations are carried out using a polynomial order of $P=4$ with $N=35937$ elements. Assuming $PPW=4$, the upper frequency is given by $f_u=2.8$ kHz. The model is excited with a Gaussian pulse as an initial condition with $\sigma_g=0.1$ m$^2$ placed at ($s_x, s_y, s_z)_{CB}=(0.7, 0.5, 0.5)$ m. The receiver position is placed at ($r_x, r_y, r_z)=(0.25, 0.25, 0.50)$ m. A second 3D room is considered to follow a good ratio (GR) of 1.9:1.4:1, which assures an even distribution of the room modes\citep{GRCH}. The room size is $(L_x,L_y,L_z)=(1.615, 1.190, 0.850)$ m, sound source is placed at ($s_x, s_y, s_z)_{GR}=(1.200, 0.600, 0.425)$ m and the receiver ($r_x, r_y, r_z)=(0.500, 0.200, 0.425)$ m. Again, a polynomial order of $P=4$ with $N=35937$ is considered. Assuming a spatial resolution corresponding to about 4 $PPW$ for the highest frequencies ($f_u=1.7$ kHz).

\subsection{Inhomogeneous boundary parameterization}
\begin{figure}
\figcolumn{%
\fig{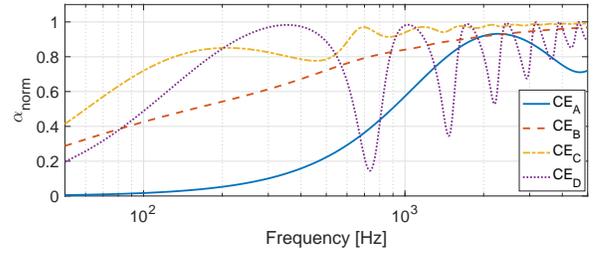}{0.5\textwidth}{(a)}\label{fig:freq_indpSOL}
\fig{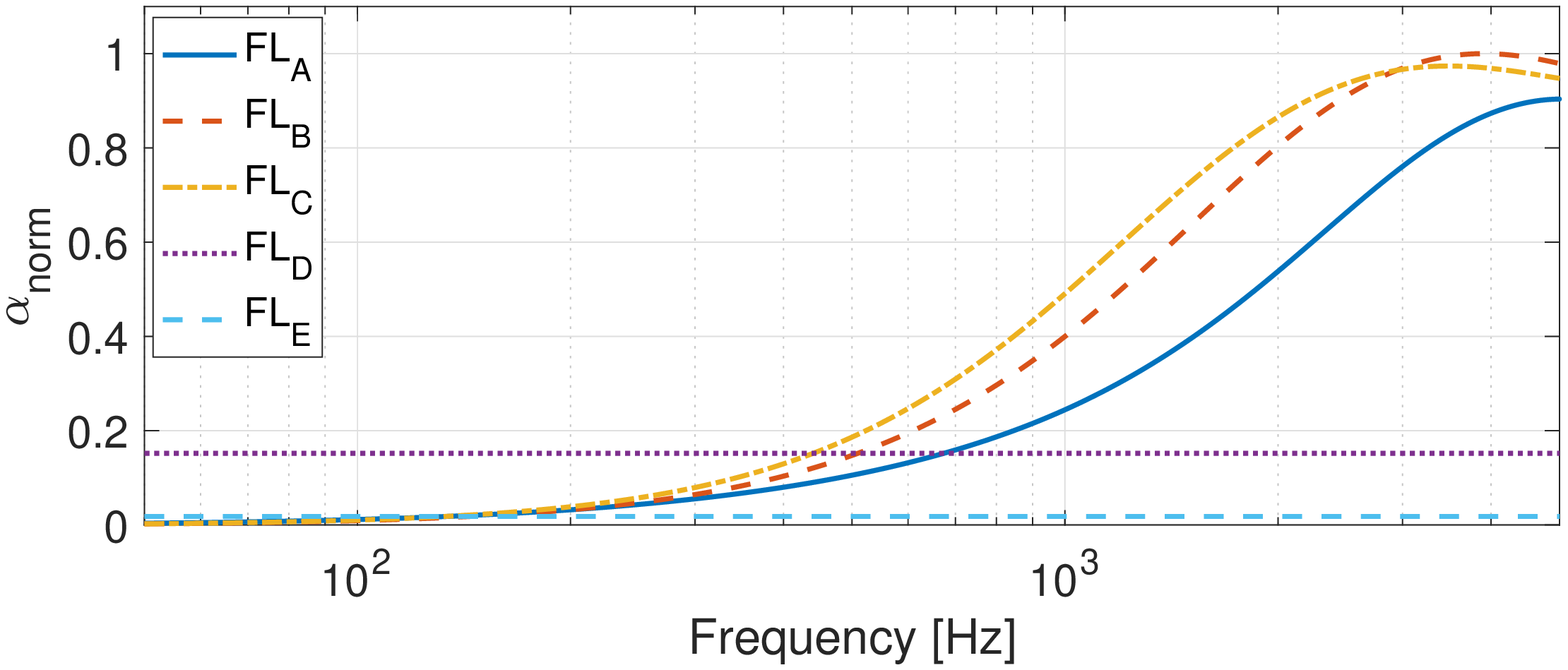}{0.5\textwidth}{(b)} \label{fig:freq_dpSOL}
\fig{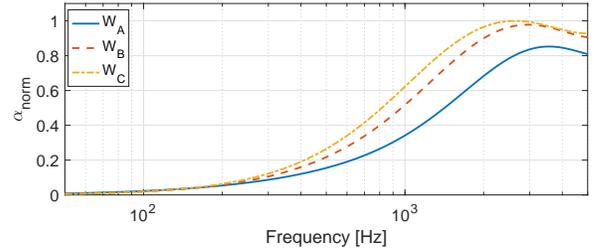}{0.5\textwidth}{(c)}}\label{fig:abscoef1}
\caption{Absorption coefficients. a) Ceiling (CE), b) Floor (FL), c) Walls (W).}
\label{fig:abscoeff}
\end{figure}

First, the 2D domain is considered in section \ref{sec:param2d}. The ceiling (CE) is modelled with a porous absorber. The key parameters affecting the absorption characteristic of a porous absorber are the flow resistivity, thickness, and air cavity depth \citep{cox}. To parameterize the porous ceiling, FOMs with $\sigma_{mat}=[10, 30, 50]$ kNsm$^{-4}$,  $d_{mat}=[0.02, 0.12, 0.22]$ m, and $d_0=[0.02, 0.12, 0.22]$ m are simulated in the offline stage. The floor (FL) is designed with two different options: as a hard surface modelled as a frequency independent boundary and covered with a carpet modelled as a porous layer. Frequency independent boundary conditions are ranges $Z_s=[10, 50, 90]$ kNsm$^{-3}$, while frequency dependent boundary conditions to model the carpet are computed with a fixed thickness of $0.02$ m but varying $\sigma_{mat}=[10, 30, 50]$ kNsm$^{-4}$. Finally, the left wall (WL) and right (WR) walls are modelled as porous panels where $d_{mat}=0.03$ m, $d_0=0$ m and $\sigma_{mat}= [5, 12, 19]$ kNsm$^{-4}$. Figure \ref{fig:abscoeff} shows the absorption coefficients for the most and least absorptive cases of each surface, whose corresponding parameter values are given in Table \ref{Tab:abscoeff}.
 
\begin{table}[]
\caption{Parameter values of the presented absorption coefficients of Figure \ref{fig:abscoeff}.} 
\label{Tab:abscoeff}
\resizebox{.35\textwidth}{!}{
\begin{tabular}{|c|c|c|c|c|}
\hline
 & \textbf{$\sigma_{mat}$}  {[}kNsm$^{-4}${]} & \textbf{$d_{mat}$}  {[}$m${]} & \textbf{$d_0$} {[}$m${]}& \textbf{$Z_s$}  {[}kNsm$^{-3}${]} \\ \hline
\textbf{CE$_A$} & 10 & 0.02 & 0.02 & - \\ \hline
\textbf{CE$_B$} & 30 & 0.12 & 0.12 & - \\ \hline
\textbf{CE$_C$} & 10 & 0.12 & 0.22 & - \\ \hline
\textbf{CE$_D$} & 30 & 0.02 & 0.22 & - \\ \hline
\textbf{FL$_A$} & 10 & 0.02 & 0 & - \\ \hline
\textbf{FL$_B$} & 30 & 0.02 & 0 & - \\ \hline
\textbf{FL$_C$} & 50 & 0.02 & 0 & - \\ \hline
\textbf{FL$_D$} & - & - & - & 10 \\ \hline
\textbf{FL$_E$} & - & - & - & 90 \\ \hline
\textbf{W$_A$} & 5 & 0.03 & 0 & - \\ \hline
\textbf{W$_B$} & 12 & 0.03 & 0 & - \\ \hline
\textbf{W$_C$} & 19 & 0.03 & 0 & - \\ \hline
\end{tabular}
}
\end{table}
 
A way to construct the ROM is by performing FOM simulations for each combination of the parameter values. To cover the inhomogeneous boundary variation, a total of $3^7$ $(2187)$ FOM simulations are possible. This is way too many for practical runtime constraints. Instead, we have chosen only $3\times 7$ $(=21)$ FOM simulations, which is shown in the Appendix (Table \ref{Tab:ROM2dparam}) indicating for each FOM simulation which parameter values are chosen and which are fixed. The table rows correspond to the 21 simulations, and the columns correspond to the different parameter values shown in Table \ref{Tab:ROM2d}. The chosen parameters are marked with an X. To the author's knowledge, this study is the first to report MOR performances with such complicated room acoustic setups and is, therefore, useful to establish the feasibility of the approach.\\

\begin{table}[]

\caption{Parameter values chosen to construct the ROM in 2D. The parameters under variation that are included in the ROM are marked with *.} 
\label{Tab:ROM2d}
\resizebox{.5\textwidth}{!}{%
\begin{tabular}{|c|c|c|c|c|}
\hline
 & \textbf{CE} & \textbf{FL} & \textbf{WL} & \textbf{WR}\\ \hline
\textbf{$\sigma_{mat}$} ${[}kNsm^{-4}${]}  & [10, 30, 50]* & [10, 30, 50]* & [5, 12, 19]* & [5, 12, 19]* \\ \hline
\textbf{$d_{mat}$} {[}$m${]} & [0.02, 0.12, 0.22]* & 0.02 & 0.03 & 0.03 \\ \hline
\textbf{$d_0$} {[}$m${]}  &  [0.02, 0.12, 0.22]* & 0 & 0 & 0 \\ \hline
\textbf{$Z_s$}  {[}kNsm$^{-3}${]} & -& [10, 50, 90]*  & - & - \\ \hline

\end{tabular}
}

\end{table}

Second, section \ref{sec:param3d} considers the 3D domain for both cube and good ratio shapes. In both cases, the ceiling (CE) is modelled as a porous acoustic material of a fixed thickness $d_{mat}=0.05$ m, where $\sigma_{mat}=[10, 30, 50]$ kNsm$^{-4}$ and $d_0=[0.02, 0.12, 0.22]$ m are parameterized. The floor (FL) is modelled as a frequency independent boundary with $Z_s=50$ kNsm$^{-3}$. The east wall (WE) is modelled as a frequency independent brick surface with $Z_s=50$ kNsm$^{-3}$. The south wall (WS) is not parameterized, and it is covered with a porous material where $\sigma_{mat}=7$ kNsm$^{-4}$, $d_{mat}=0.02$ m and $d_0=0$ m. The west wall (WW) is modelled with a porous material of $d_{mat}=0.05$ m and $d_0=0$ m whose flow resistivity is parameterized with values of $\sigma_{mat}=[5, 12, 19]$ kNsm$^{-4}$. The north wall (WN) is modelled as a hard surface with $Z_s=50$ kNsm$^{-3}$. In addition, a $0.5\times 0.5$ m$^2$ square acoustic panel made of porous material is placed at the centre of the wall. The panel has a fixed thickness and flow resistivity of $d_{mat}=0.1$ m and $\sigma_{mat}=30$ kNsm$^{-4}$ respectively. The air gap between the panel and the wall is parameterized $d_0=[0.2, 0.12, 0.22]$ m. Table \ref{Tab:ROM3d} summarizes the parameter values of each surface.\\

The ROM is constructed for each 3D room constructed with four different parameters (2 for CE, 1 for WW and 1 for WN) with three different values each. Thus, a total number of $3^4=81$ FOM simulations are possible. Instead, $3\times 4=12$ FOM simulations were carried out to construct the ROM in the same way described for the 2D case. The chosen parameter values for each FOM simulation are shown in the Appendix (Table \ref{Tab:ROM3dparam}). \\

\begin{table}[]

\caption{Parameter values chosen to construct the ROM in 3D. The parameters under variation that are included in the ROM are marked with *.} 
\label{Tab:ROM3d}
\resizebox{.5\textwidth}{!}{%
\begin{tabular}{|c|c|c|c|c|c|c|}
\hline
 & \textbf{CE} & \textbf{FL} & \textbf{WE} & \textbf{WS} & \textbf{WW}  & \textbf{WN} \\ \hline
\textbf{$\sigma_{mat}$} {[}$kNsm^{-4}${]}  & [10, 30, 50]* & -& - &70&[5, 12, 19]*&30 \\ \hline
\textbf{$d_{mat}$} {[}$m${]} & 0.05 & - & -&0.02&0.05&0.1 \\ \hline
\textbf{$d_0$} {[}$m${]}  &  [0.02, 0.12, 0.22]* & - & - & 0 &0& [0.02, 0.12, 0.22]* \\ \hline
\textbf{$Z_s$} {[}$kNsm^{-3}${]} & -& 50 & 50 & -&-&- \\ \hline

\end{tabular}
}

\end{table}

\section{\label{sec:3} Results}
This section presents results mostly with the singular energy decay, E/Eo, relative error shown in (\ref{eq:errel}), speedups (\ref{eqspeedup}), sound pressure level (SPL) spectrum, and the reverberation time (RT).
\subsection{2D - Influence of the source position and geometry}\label{sec:influece}
 
\begin{figure}
\fig{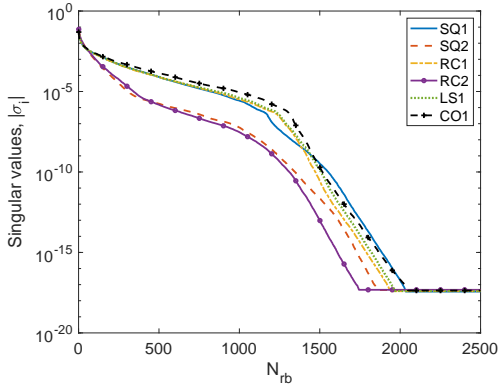}{7cm}{}
\caption{Singular value decay for the first 2500 modes of the basis and energy distribution $(E/E_0)$ among the basis with frequency independent boundaries.}
\label{fig1}
\end{figure}

For the geometries and source positions tested, the singular value decay is shown in Figure \ref{fig1}. A faster decay means that a  larger portion of the energy is concentrated in the first singular values, effectively indicating that a smaller number of basis $N_{rb}$ is needed for a given error tolerance. Slower decays would need a larger number of $N_{rb}$ to provide the same error. Note that the smaller $N_{rb}$ is, the higher speedups are achieved. In Figure \ref{fig1}, one can see that the more symmetric the problem is, both in terms of geometry and source location, the faster the decay is, as measured in the singular values translating into more efficiency (speedup). For example, $SQ$ shows a faster decay than $RC$ and $LS$. However, the difference is not significant until the energy reaches the value of $10^{-10}$, and it can be concluded that the room geometry does not significantly change the energy distribution among the basis. On the other hand, the centred sound source locations lead to a faster decay of the singular values compared to the corner source. $SQ2$ shows a faster decay than $SQ1$, and the same for $RC2$ in comparison to $RC1$. This is because placing a source at the centre of the room will fail to excite some room modes, of which the nodal lines/points coincide with the source location. This leads to a smaller number of basis needed to describe the physical dynamics of the wave propagation in the room accurately. 



\subsection{2D - Parameterization of different absorption properties}\label{sec:param2d}
\begin{figure}
\fig{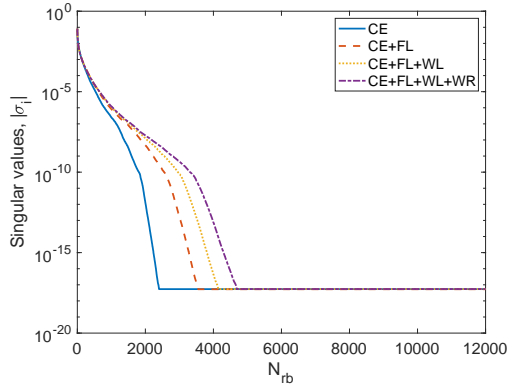}{7cm}{}
\caption{Singular value decay for different number of parameters for case 1 and case 2.}
\label{fig:multidecay}
\end{figure}

In Figure \ref{fig:multidecay}, the singular value decay when a different number of parameters are included in the model, i.e., only the ceiling parameters are included in the ROM (CE); the ceiling and the floor parameters (CE+FL); the ceiling, the floor and the left wall (CE+FL+WL) and all the sides (CE+FL+WL+WR). The more parameters added, the slower the decay curve, so adding more parameters to the ROM has a clear effect on the singular value decay and, thus, the choice of the basis for ROM. However, the decay remains similar in the first basis functions, which are the ones used to construct the ROM. \\

\begin{table}[]
\caption{Parameter values for the online stage of the 2D ROM. Values marked with * denotes the parameters which are parameterized.} 
\label{Tab:cases}
\resizebox{.35\textwidth}{!}{
\begin{tabular}{|c|c|c|c|c|}
\hline
 & \textbf{$\sigma_{mat}$} {[}$kNsm^{-4}${]} & \textbf{$d_{mat}$} {[}$m${]} & \textbf{$d_0$} {[}$m${]} & \textbf{$Z_s$}  {[}$kNsm^{-3}${]}\\ \hline
\textbf{CE1} & 2* & 0.1* & 0.1* & - \\ \hline
\textbf{FL1} & 12* & 0.02 & 0 & - \\ \hline
\textbf{WL1} & 10* & 0.03* & 0 & - \\ \hline
\textbf{WR1} & 15* & 0.03 & 0 & - \\ \hline \hline
\textbf{CE2} & 45* & 0.05* & 0.2* & - \\ \hline
\textbf{FL2} & - & - & - & 7* \\ \hline
\textbf{WL2} & 10* & 0.2* & 0 & - \\ \hline
\textbf{WR2} & 6* & - & 0 & - \\ \hline
\end{tabular}
}
\end{table}

\begin{figure}
\figcolumn{%
\fig{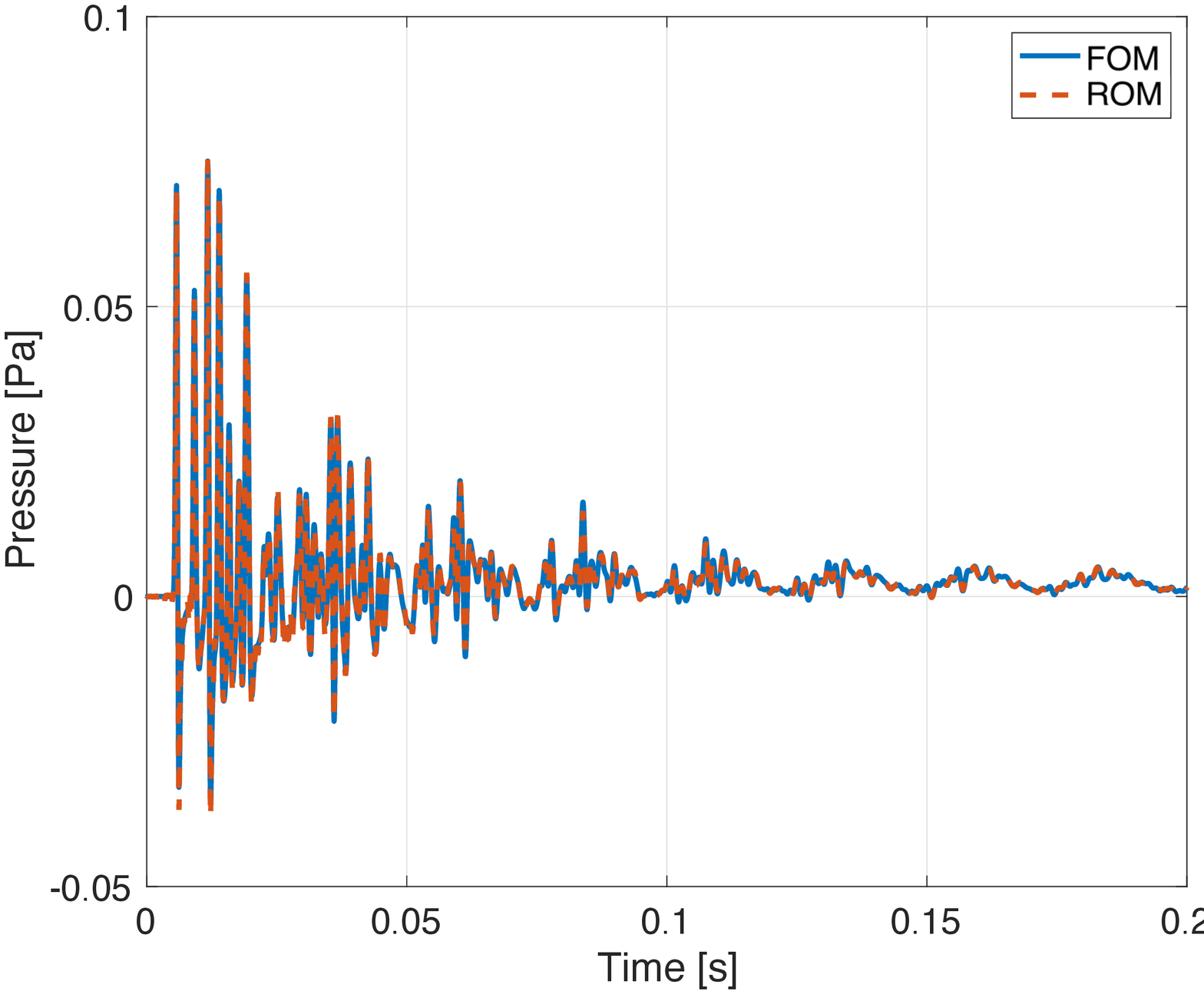}{.4\textwidth}{(a)}\label{fig:res2dtime}
\fig{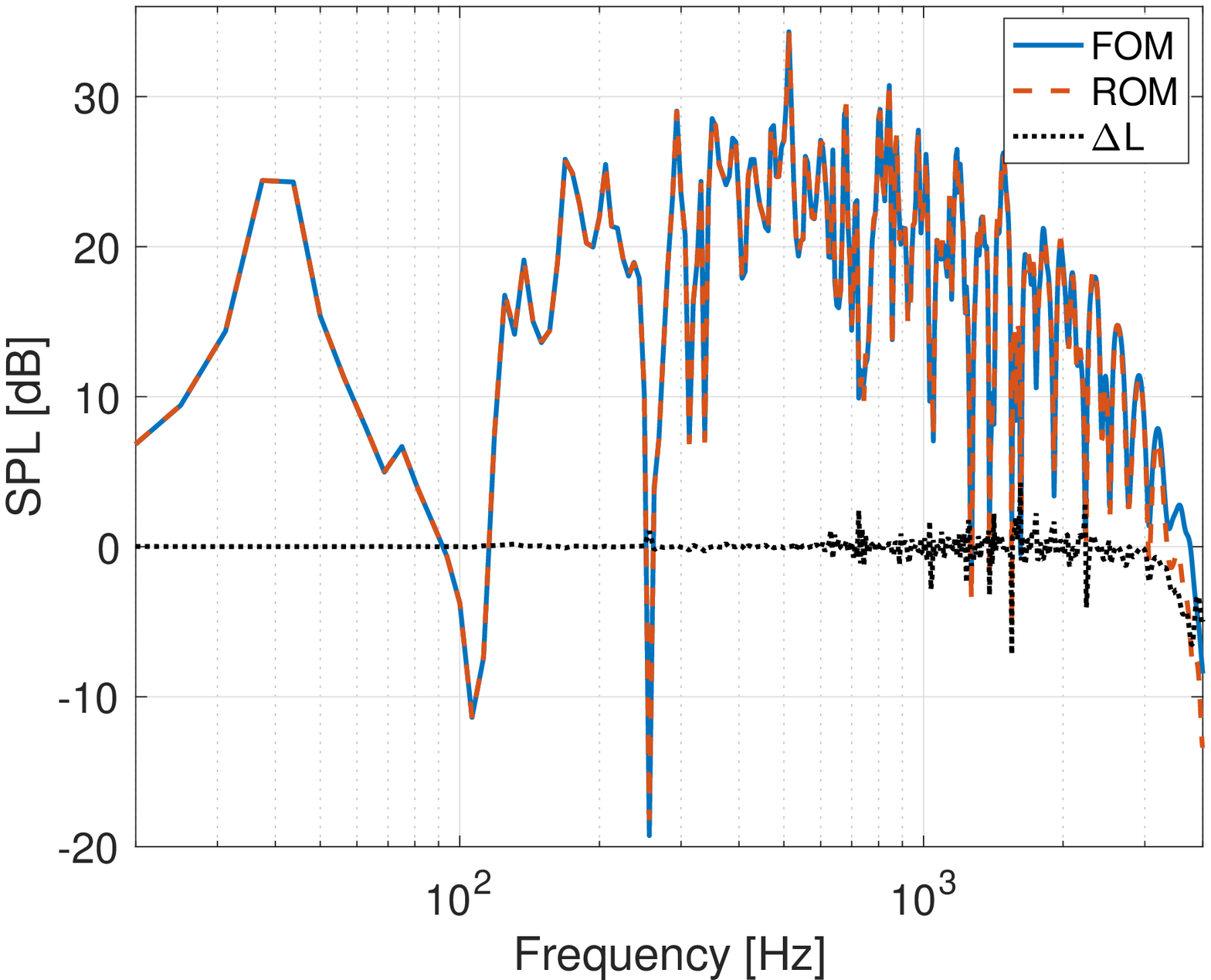}{.4\textwidth}{(b)}}\label{fig:res2dfreq4k}
\caption{(a) Impulse response and (b) spectrum of 2D FOM and ROM of case 1 for $N_{rb}=531$ and $f_u=4$ kHz (using the parameter in Table \ref{fig:multidecay}).}
\label{fig:res2d}
\end{figure}

\begin{figure}
\fig{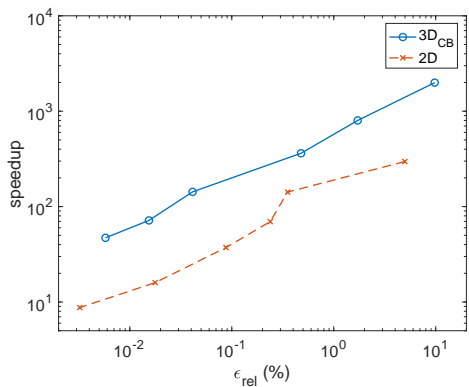}{7cm}{}
\caption{Speedup against the relative rms error for the 2D domain (case 1) with $f_u=2$ kHz, and 3D CB.}
\label{fig:speedup}
\end{figure}

In the online stage, two ROM cases are simulated and compared with their corresponding FOM. Case 1 deals with frequency dependent boundaries, while case 2 allows to change the floor to a rigid surface, modelled as a frequency independent boundary. The boundary parameter values are shown in Table \ref{Tab:cases}, and the IR and SPL are shown in Figure \ref{fig:res2d}. Several different online ROM simulations are compared for $N_{rb}= 69, 158, 274, 410, 571,752$ for the ROM with an upper frequency of 2 kHz corresponding to values of $\epsilon_{POD}=10^{-2},10^{-3},...,10^{-7}$; and $N_{rb}= 132, 310, 531, 810$ for the ROM with an upper frequency of 4 kHz corresponding to values $\epsilon_{POD}=10^{-2},...,10^{-5}$. Figure \ref{fig:res2d} shows the results up to 4 kHz with $N_{rb}=531$. Figure \ref{fig:res2dtime} shows the impulse response, and Figure \ref{fig:res2dfreq4k} presents the sound pressure level (SPL) from 20 Hz to 4 kHz showing that $\Delta L$ is nearly zero below $600$ Hz and increasing with the frequency. Note that around $4$ kHz, a roll-off is presented due to the source excitation. This case enables a computational speedup by a factor of 37 for an error of $\epsilon_{rel}=0.03\%$. For the ROM constructed with $N_{rb}=274$ and an upper frequency of 2 kHz, the error is $\epsilon_{rel}=0.2\%$ for case 1 with a speedup of 70. Moreover, case 2 presents an error of $\epsilon_{rel}=0.3\%$ and a speedup of 50. Note that the accuracy depends on $N_{rb}$. The speedup against  $\epsilon_{rel}$ given in (\ref{eq:errel}) is shown in Figure \ref{fig:speedup}. These results show speedups around two orders of magnitude for 2D. 

\subsection{3D - Parameterization of different absorption properties}\label{sec:param3d}
\begin{figure}
\figcolumn{%
\fig{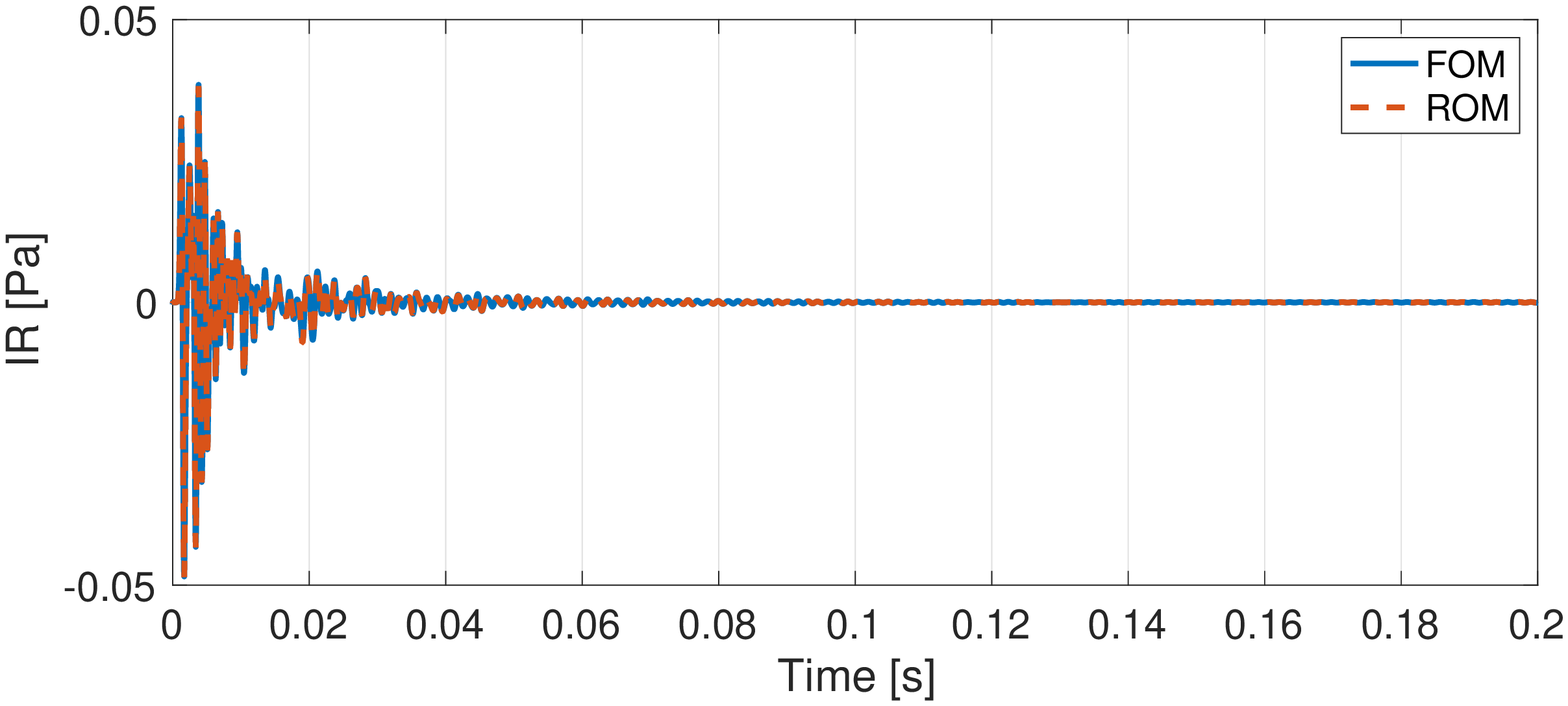}{.5\textwidth}{(a)}\label{fig:res3dtime}
\fig{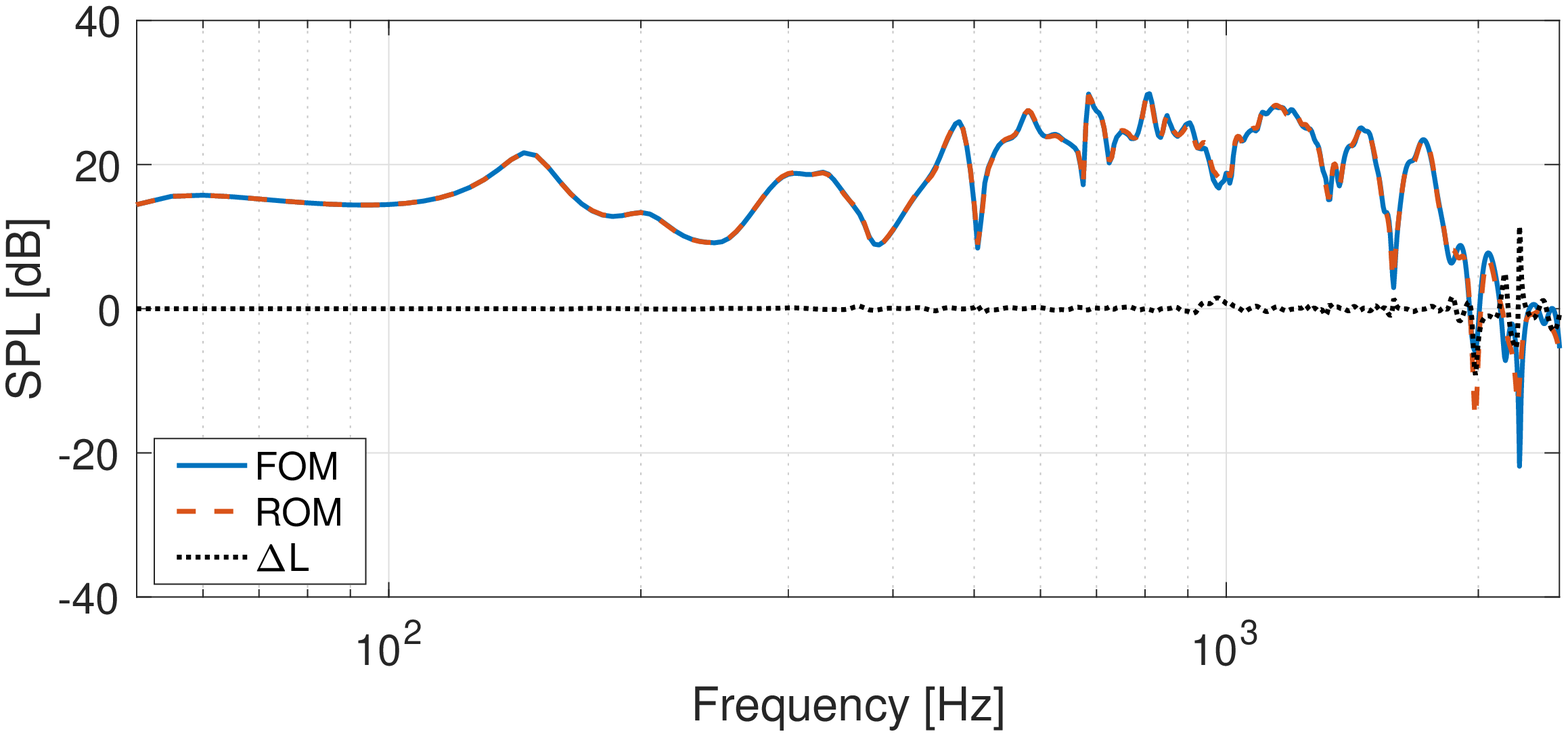}{.5\textwidth}{(b)}\label{fig:res3dfreq}
\fig{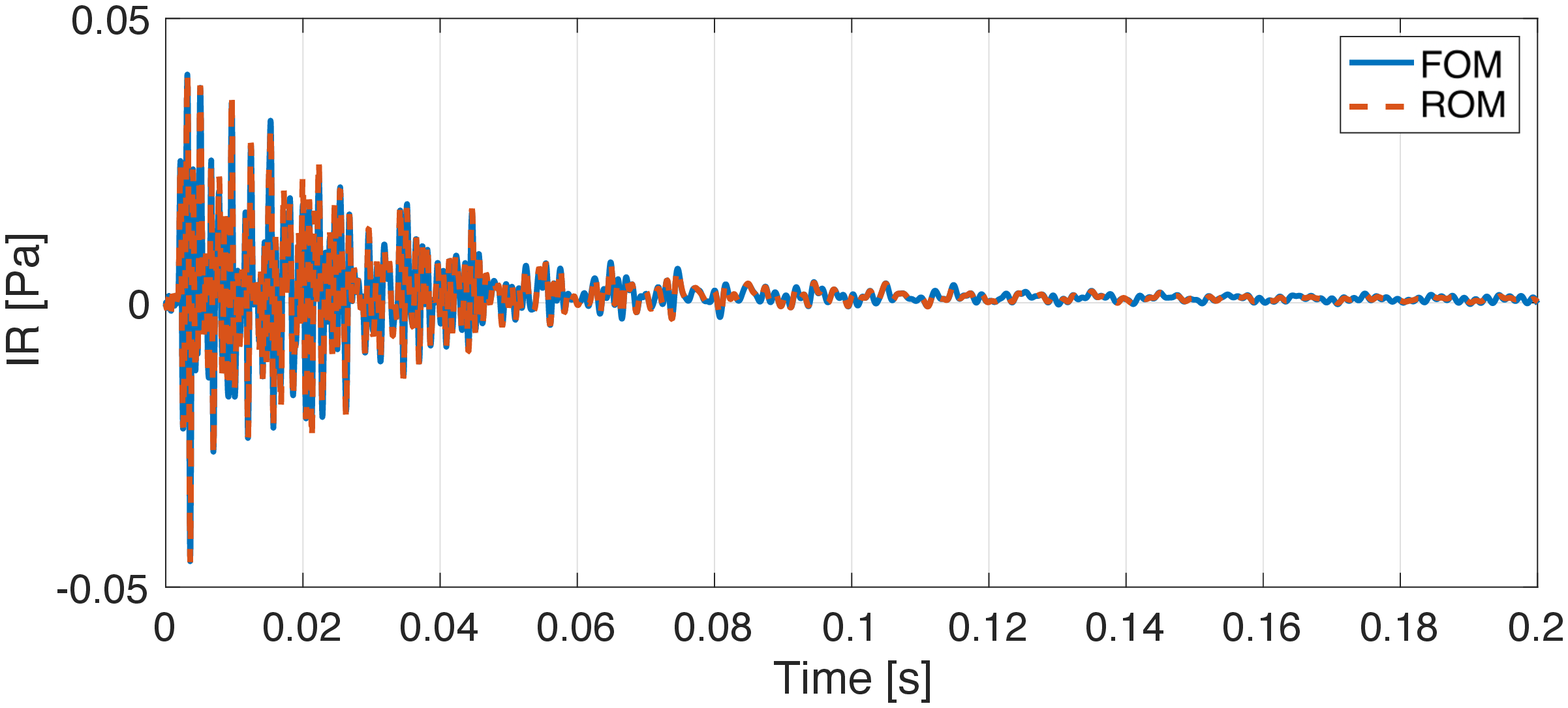}{.5\textwidth}{(c)}\label{fig:res3dtimeGR}
\fig{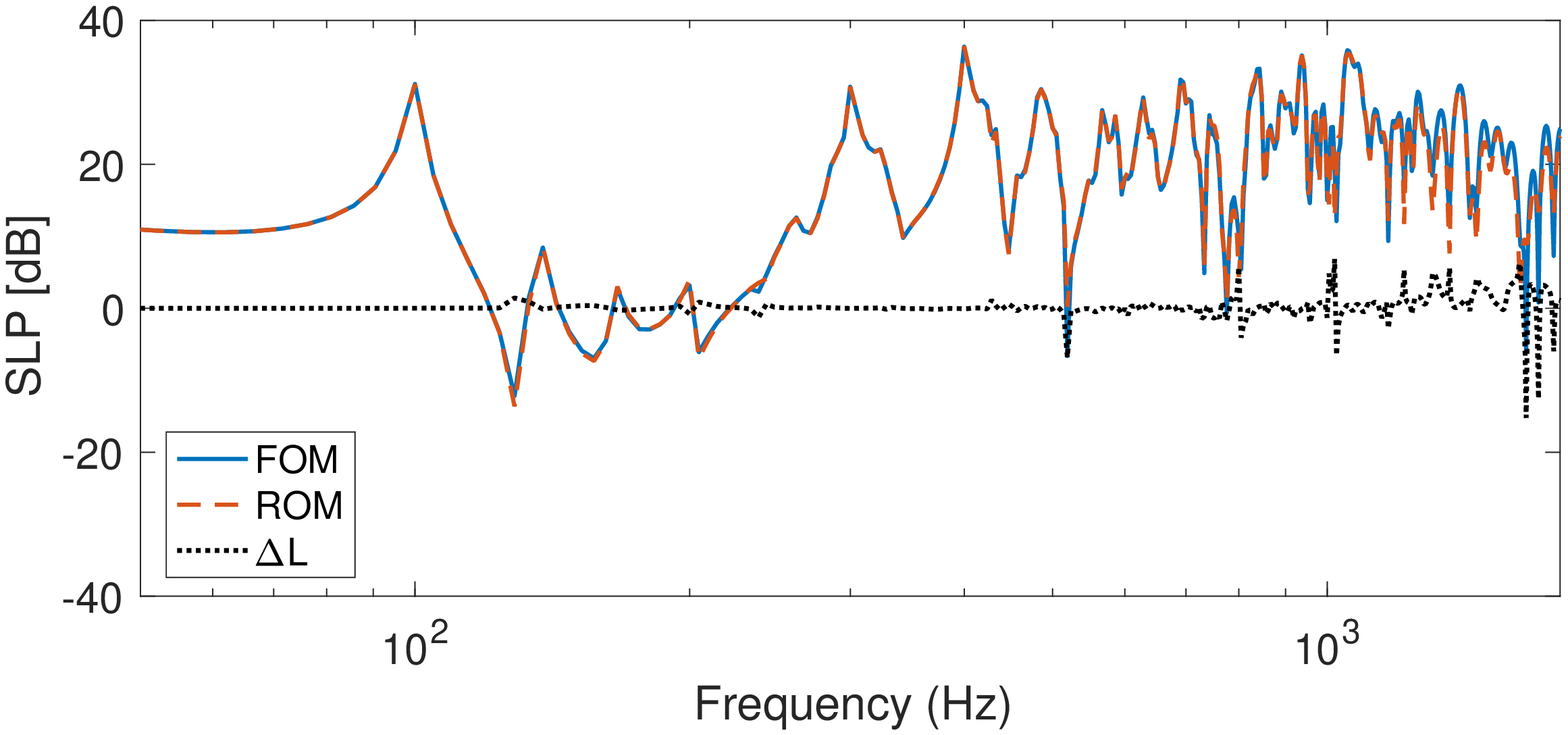}{.5\textwidth}{(d)}\label{fig:res3dfreqGR}}
\caption{Simulated pressure using the 3D FOM and ROM. a) CB domain sound pressure  with $N_{rb}=275$, b) CB domain frequency response  with $N_{rb}=275$, c) GR domain sound pressure  with $N_{rb}=478$, d) GR domain frequency response  with $N_{rb}=478$.  }
\label{fig:res3d}
\end{figure}

\begin{figure}
\figcolumn{%
\fig{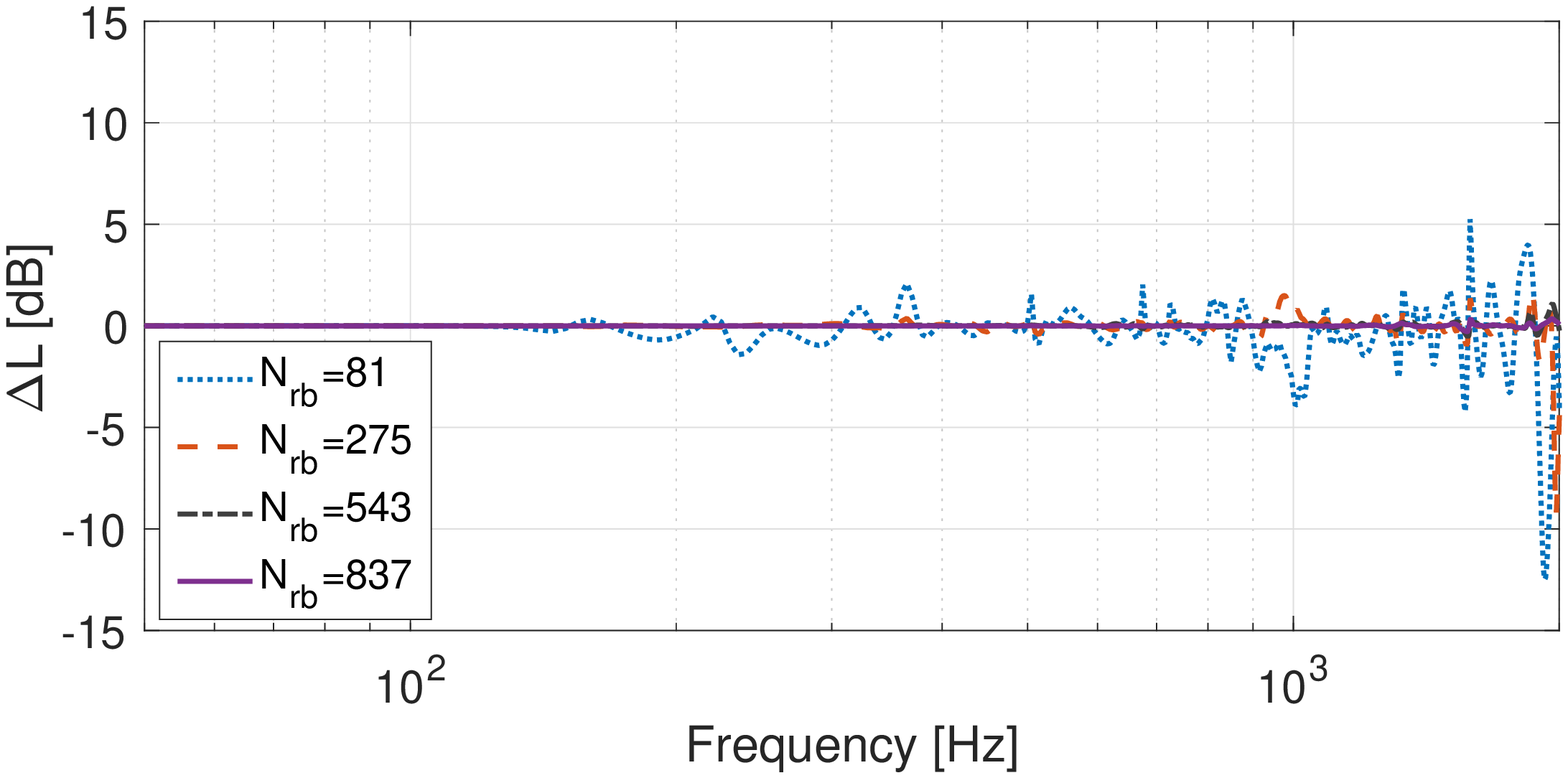}{.5\textwidth}{(a)}\label{fig:deltaerrorCB}
\fig{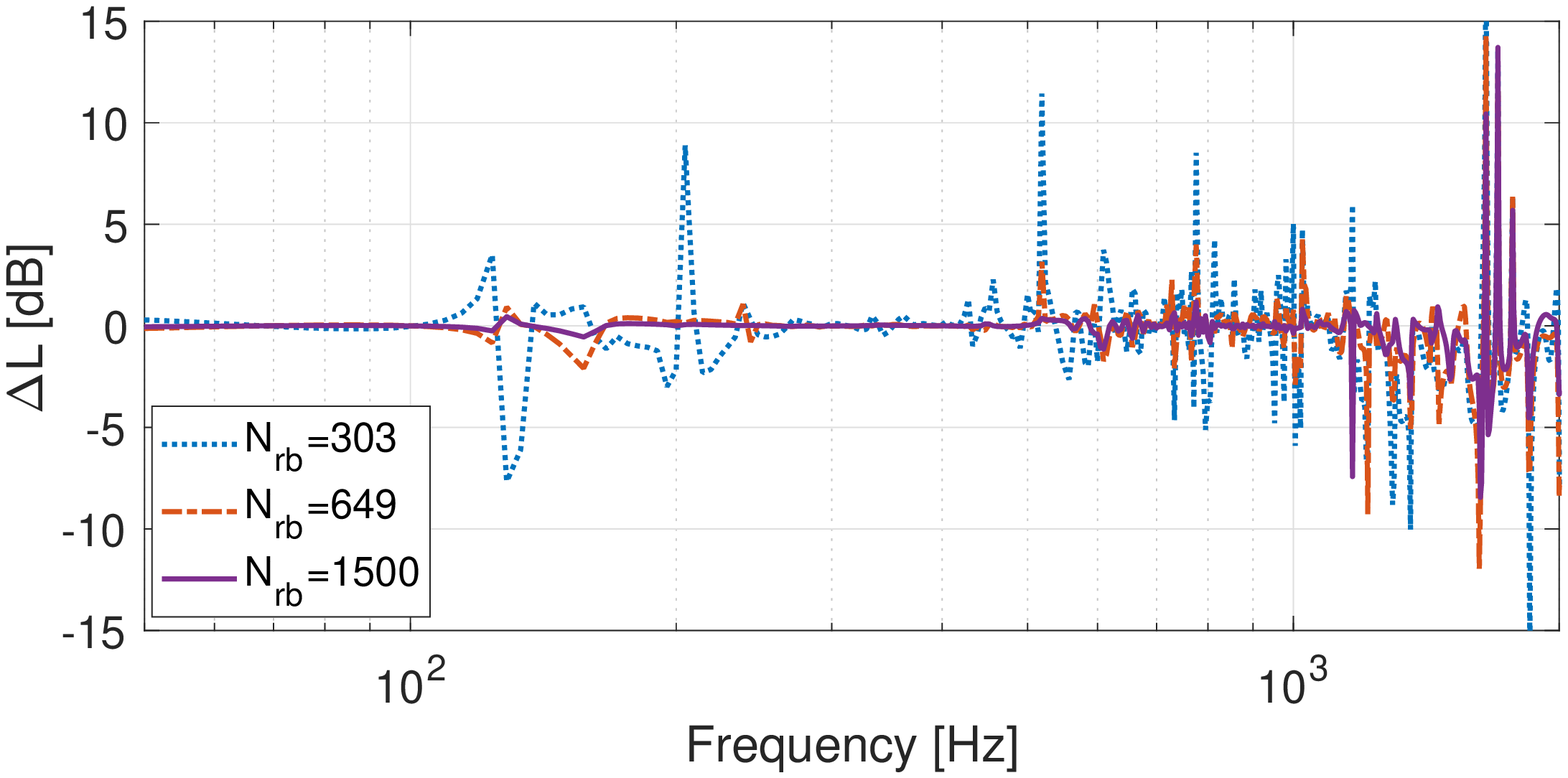}{.5\textwidth}{(b)}\label{fig:deltaerrorGR}}
\caption{$\Delta L$ error. a) CB case 1 for $N_{rb}= 81, 275, 543,837$. b) GR for $N_{rb}= 303, 649, 1500$. }
\label{fig:erroroctbands}
\end{figure}

This section presents a similar analysis with 3D test cases. By varying $N_{rb}$, simulations are compared. The chosen parameters for the online simulation are shown in Table \ref{Tab:cases3D}. For the CB the following number of basis are considered $N_{rb}= 30, 81,175,275,543,837$ corresponding to values of $\epsilon_{POD}=10^{-2},...,10^{-9}$ to be compared against the corresponding FOM solution for verification purposes. Moreover, for the GR room, the ROM was constructed with $N_{rb}= 62, 303, 478, 649, 1500$ corresponding to values of $\epsilon_{POD}=10^{-2},...,10^{-9}$. Table \ref{Tab:cases3D} shows the chosen parameter values for all the surfaces for both the cubic and good ratio domains. Figure \ref{fig:res3dtime} and Figure \ref{fig:res3dfreq} present the impulse response and frequency response, respectively, for the CB. Moreover,  Figure \ref{fig:res3dtimeGR} and Figure \ref{fig:res3dfreqGR} show the impulse response and frequency response, respectively, for the GR room. In both cases, the $\Delta L$ is included, which confirms a good agreement between FOM and ROM for the given $N_{rb}$. The CB is constructed with $N_{rb}=273$ where the error is $\epsilon_{rel}=0.04\%$ with a speedup of 143. On the other hand, the GR room is constructed with $N_{rb}=478$, where the error is $\epsilon_{rel}=0.66\%$ with a speedup of 90. Note $\Delta L$ in Eq. (\ref{eqerrorL}) and  $\epsilon_{rel}$ are also presented for the different number of basis $N_{rb}$ in Figure \ref{fig:erroroctbands} for the CB (Figure \ref{fig:deltaerrorCB}) and the good ratio room (Figure \ref{fig:deltaerrorGR}). Again, the $\Delta L$ is nearly zero at lower frequencies and increases with frequency, showing more differences at the anti-resonances. The speedup against the error for CB case 1 is presented in Figure \ref{fig:speedup}, showing values around three orders of magnitude. 

\begin{figure}
\fig{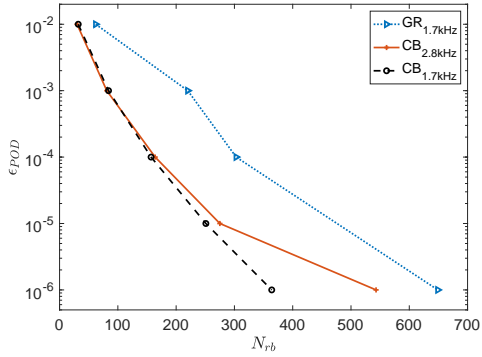}{7cm}{}
\caption{Number of basis functions $N_{rb}$ against the tolerance $\epsilon_{POD}$ introduced in (\ref{romNrbselec}) for GR and CB at two different upper frequencies $1.7$ kHz and $2.8$ kHz}
\label{fig:cbvrgr}
\end{figure}

In order to understand the behaviour of different room ratios, the number of basis functions $N_{rb}$ for a given tolerance $\epsilon_{POD}$ described in (\ref{romNrbselec}) is compared. A new ROM of the cubic room is computed with $f_u=1.7$ kHz. Figure \ref{fig:cbvrgr} shows the comparison for GR$_{1.7kHz}$, CB$_{1.7kHz}$ and CB$_{2.8kHz}$. Results show that for a fixed upper frequency, a less symmetric geometry GR$_{1.7kHz}$ needs more basis functions for a given $\epsilon_{POD}$ compared to a symmetric geometry CB$_{1.7kHz}$. This would not necessarily lead to lower speedups considering that GR has a larger number of DOF. Moreover, comparing the curves corresponding to GR$_{1.7kHz}$ and CB$_{2.8kHz}$ for a fixed number of DOF, GR$_{1.7kHz}$ results in a larger number of $N_{rb}$ for a given $\epsilon_{POD}$. Thus, it will lead to lower speedups as the numerator of equation (\ref{eqspeedup}) remains the same for GR$_{1.7kHz}$ and CB$_{2.8kHz}$. At the same time, the denominator becomes larger for GR$_{1.7kHz}$ at a given $\epsilon_{POD}$ compared to CB$_{2.8kHz}$.

\begin{figure}
 \centering
\figcolumn{\fig{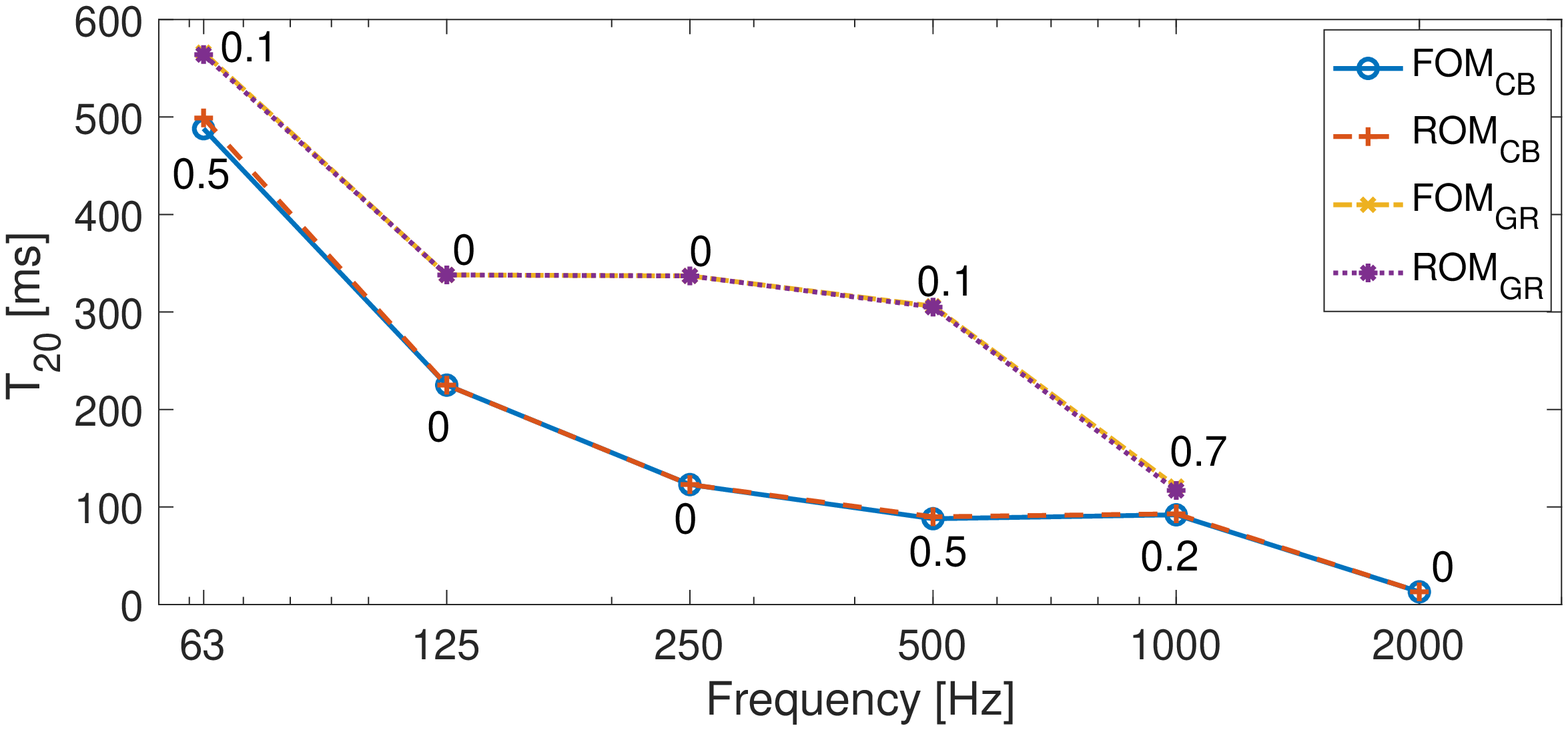}{9cm}{(a)}\label{fig:reverberation3da}
\fig{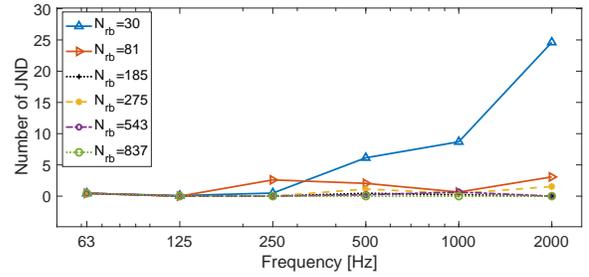}{9cm}{(b)}\label{fig:reverberation3b}
\fig{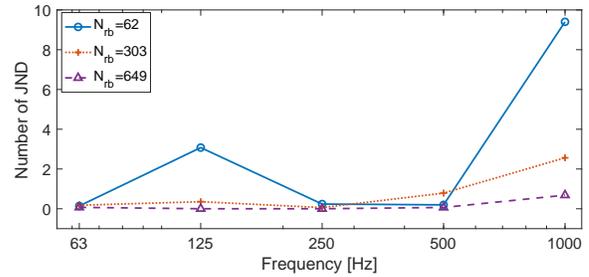}{9cm}{(c)}}\label{fig:reverberation3c}
\caption{Comparison of reverberation time between FOM and ROM and their difference in terms of numbers of JND. a) Cube domain with $N_{rb}=185$ (CB) and  good ratio domain with $N_{rb}=649$ (GR) including the number of JNDs per octave band, b) Number of JNDs for CB, c) Number of JNDs for GR.}
\label{fig:reverberation3d}
\end{figure}

The reverberation time is one of the most widely used acoustic parameters defined in ISO 3382-1 \citep{ISO3382-1} as the time needed for the energy to decrease by 60 dB. A way to evaluate the error of the reverberation time between the FOM and ROM is by means of the JND, which is the minimum change in the RT that can be perceptually perceived. The JND of the RT is 5\% as defined in the standard \citep{ISO3382-1}. Note that if the difference is larger than 1 JND, the IR can be potentially differently heard. $T_{20}$ is calculated from IRs via FOM and ROM to quantify how ROM degrades the accuracy of RT. Figure \ref{fig:reverberation3da} shows the RT for different frequency octave bands. The CB ROM is performed with $N_{rb}=185$ while the GR ROM with $N_{rb}=649$. The RT difference is below one JND in all the frequency octave bands in both cases, which indicates that the present ROM would not be perceptually different compared to the FOM. Moreover, Figure \ref{fig:reverberation3b} and Figure \ref{fig:reverberation3c} show the numbers of JND for $T_{20}$ for various $N_{rb}$. Decreasing $N_{rb}$ increases the RT difference in the higher frequency bands. Note that in this case the ROM is 365 times faster than the FOM and the CB needs fewer basis functions than GR, which supports the finding that symmetric conditions are more favourable for higher reductions.\\


\begin{table}[]
\caption{Online stage boundary parameters for the 3D rooms. Values marked with * denotes parameterization.} 
\label{Tab:cases3D}
\resizebox{.35\textwidth}{!}{
\begin{tabular}{|c||cccccc|}
\cline{1-7}
 \multicolumn{7}{|c|}{CB domain}\\ \hline
 & \multicolumn{1}{c|}{\textbf{CE}} & \multicolumn{1}{c|}{\textbf{FL}} & \multicolumn{1}{c|}{\textbf{WE}} &
\multicolumn{1}{c|}{\textbf{WS}}&
\multicolumn{1}{c|}{\textbf{WW}} &
\multicolumn{1}{c|}{\textbf{WN}} \\ \hline \hline

\multicolumn{1}{|c||}{\boldsymbol{$\sigma_{mat}$} {[}$kNsm^{-4}${]}}   & \multicolumn{1}{c|}{12*} & \multicolumn{1}{c|}{-} & \multicolumn{1}{c|}{-} & \multicolumn{1}{c|}{7} & \multicolumn{1}{c|}{10*}&\multicolumn{1}{c|}{30} \\ \hline

\multicolumn{1}{|c||}{\boldsymbol{$d_{mat}$} [m]} & \multicolumn{1}{c|}{0.05} & \multicolumn{1}{c|}{-} & \multicolumn{1}{c|}{-} &\multicolumn{1}{c|}{0.02} &\multicolumn{1}{c|}{0.05} &\multicolumn{1}{c|}{0.1}  \\ \hline

\multicolumn{1}{|c||}{\boldsymbol{$d_0$} [m]} & \multicolumn{1}{c|}{0.06*} & \multicolumn{1}{c|}{-} & \multicolumn{1}{c|}{-} & \multicolumn{1}{c|}{0} &\multicolumn{1}{c|}{0} &\multicolumn{1}{c|}{0.1*} \\ \hline

\multicolumn{1}{|c||}{\boldsymbol{$Z_s$}  {[}$kNsm^{-3}${]}} & \multicolumn{1}{c|}{-} & \multicolumn{1}{c|}{50} & \multicolumn{1}{c|}{50} & \multicolumn{1}{c|}{-} & \multicolumn{1}{c|}{-}& \multicolumn{1}{c|}{50} \\ \hline

\hline \hline 
 \multicolumn{7}{|c|}{GR domain}\\ \hline
& \multicolumn{1}{c|}{\textbf{CE}} & \multicolumn{1}{c|}{\textbf{FL}} & \multicolumn{1}{c|}{\textbf{WE}} &
\multicolumn{1}{c|}{\textbf{WS}}&
\multicolumn{1}{c|}{\textbf{WW}} &
\multicolumn{1}{c|}{\textbf{WN}} \\ \hline \hline

\multicolumn{1}{|c||}{\boldsymbol{$\sigma_{mat}$} {[}$kNsm^{-4}${]}}   & \multicolumn{1}{c|}{10.5*} & \multicolumn{1}{c|}{-} & \multicolumn{1}{c|}{-} & \multicolumn{1}{c|}{7} & \multicolumn{1}{c|}{5.5*}&\multicolumn{1}{c|}{30} \\ \hline

\multicolumn{1}{|c||}{\boldsymbol{$d_{mat}$} [m]} & \multicolumn{1}{c|}{0.05} & \multicolumn{1}{c|}{-} & \multicolumn{1}{c|}{-} &\multicolumn{1}{c|}{0.02} &\multicolumn{1}{c|}{0.05} &\multicolumn{1}{c|}{0.1}  \\ \hline

\multicolumn{1}{|c||}{\boldsymbol{$d_0$} [m]} & \multicolumn{1}{c|}{0.025*} & \multicolumn{1}{c|}{-} & \multicolumn{1}{c|}{-} & \multicolumn{1}{c|}{0} &\multicolumn{1}{c|}{0} &\multicolumn{1}{c|}{0.03*} \\ \hline

\multicolumn{1}{|c||}{\boldsymbol{$Z_s$}  {[}$kNsm^{-3}${]}} & \multicolumn{1}{c|}{-} & \multicolumn{1}{c|}{50} & \multicolumn{1}{c|}{50} & \multicolumn{1}{c|}{-} & \multicolumn{1}{c|}{-}& \multicolumn{1}{c|}{50} \\ \hline

\end{tabular}
}
\end{table}

\section{\label{sec:4}Discussion}
The performance behaviour of the ROM in terms of speedups is case-dependent and can be challenging to predict, especially when including a large number of parameters in a complex scene. This study analyzes the behaviour of realistic room acoustic scenarios by varying the geometry, source location, and inhomogeneous boundary in 2D and 3D. According to the singular energy decay, the geometry of the room does not have a significant impact on choosing the reduced basis and, therefore, the speedup.
Moreover, the performance of the ROM when adding new parameters can also be estimated based on previous calculations, as it has been shown in Figure \ref{fig:multidecay} that the singular value decay is practically the same in the first basis functions when adding new parameters to the ROM, which is a clear indication of the potential of ROM for the type of applications studied.

When increasing the dimension of a space (from 2D to 3D), a higher speedup is resulted for the same error values (Figure \ref{fig:speedup}). Speedups of up to two orders of magnitude are found for 2D and up to three orders for 3D simulations when compared to FOM. These results agree with previous studies with homogeneous boundary conditions \citep{RBMHermes}.

A recent MOR in time domain room acoustic simulations using an automated Krylov subspace algorithm reported a speedup of 11–36 without introducing audible differences for a simple scenario \citep{Miller2}. The present study shows higher speedups for a larger domain (considering the higher frequency). For the 2D domain (cases 1) with $f_u=2$ kHz, a reduction of the degree of freedom from DOF$= 12039$ to $N_{rb}= 158$ and $N_{rb}= 752$ resulted in a speedup of $142$ ($\epsilon_{rel}=0.35\%$) and $9$ ($\epsilon_{rel}=3.2\times 10^{-3}\%$) respectively. Moreover, for the 3D cubic case, a reduction of the degree of freedom from DOF$=35937$ to $N_{rb}= 81$ and $N_{rb}= 837$ resulted in a speedup of $800$ ($\epsilon_{rel}=1.7\%$) and $47$ ($\epsilon_{rel}=5.8\times 10^{-3}\%$) respectively. 

The difference in the reverberation between FOM and ROM has been quantified in relation to $5\%$ JND of RT defined in \citep{ISO3382-1}. Note that different studies show that the perception of the reverberation varies depending on the sound decay \citep{jnd6} and the nature of the stimuli \citep{jnd2,jnd4,jnd5, jnd3}, where the JND range from $3\%$-$20\%$.
 


\section{\label{sec:5}Conclusion}
This study is concerned with the ability of ROM for use with different boundary conditions under the variation of a considerable number of parameters and an inhomogeneous distribution of absorption across the different surfaces of a room. 
First, our results confirm that the RBM is more favourable in terms of computational reduction for symmetric problems, e.g., source positioned at the centre than in a corner. Second, results show that the singular value decay becomes more gentle when including more parameters into the ROM. Thirdly, speedups of one-two orders of magnitude are found for 2D, while two-three orders of magnitude are found for 3D. Fourthly, an analysis of reverberation time confirms that ROM produces IRs of which the RTs are less than $5\%$ from FOM. A smaller number of basis modes is needed in the truncated basis for the symmetric case in 3D, which shows a performance 365 times faster than the FOM.

It can be concluded that complex ROMs with a large number of parameters and with acoustic materials distributed inhomogeneously behave similarly to simple and homogeneous models and can achieve similar speedup performance. However, non-symmetric source positions and geometries and a large number of parameters can lead to a slower singular value decay, which may decrease the reduction if a large number of basis functions are included in the ROM.
Although the FOM simulations are computationally costly, the price paid to create the ROM using FOM simulations is worthy for multiple design evaluations, where different parameter configurations are to be explored or optimized for room acoustics.

%

\begin{acknowledgments}
This research was done at the Technical University of Denmark and partly supported by Innovationsfonden, Denmark
(Grant ID 9065-00115B), Rambøll Danmark A/S and Saint-Gobain Ecophon A/S, Sweden.
\end{acknowledgments}

\section*{References}
\bibliography{Manuscript.bib}

\begin{thebibliography}{10}
\def\enquote#1,{``#1,''}
\def\enxquote#1{``#1''}
\expandafter\ifx\csname url\endcsname\relax
  \def\url#1{\texttt{#1}}\fi
\expandafter\ifx\csname urlprefix\endcsname\relax\def\urlprefix{URL }\fi
\providecommand{\bibinfo}[2]{#2}
\def\plainquote#1{``#1''}
\providecommand{\noopsort}[1]{}
\providecommand{\switchargs}[2]{#2#1}
\providecommand{\dourl}[1]{\href{http://#1}{\nolinkurl{#1}}}
  \def\eatspace #1{#1}

\bibitem{officeproductivity1}
\bibinfo{author}{D.~Abbot}, \enquote{\bibinfo{title}{Calming the office
  cacophony}},  \bibinfo{journal}{Safety Health Pract.} \textbf{22}(1),
  \bibinfo{pages}{34--36} (\bibinfo{year}{2004}).

\bibitem{officeproductivity2}
\bibinfo{author}{N.~A. Oseland} and \bibinfo{author}{A.~Burton},
  \enquote{\bibinfo{title}{Quantifying the impact of environmental conditions
  on worker performance for inputting to a business case to justify enhanced
  workplace design features}},  \bibinfo{journal}{J. Build. Survey, Appr. Val.}
  \textbf{1}(2), \bibinfo{pages}{151--164} (\bibinfo{year}{2012}).

\bibitem{schools1}
\bibinfo{author}{B.~M. Shield} and \bibinfo{author}{J.~E. Dockrell},
  \enquote{\bibinfo{title}{The effects of noise on children at school: A
  review}},  \bibinfo{journal}{Build. Acoust.} \textbf{10}(2),
  \bibinfo{pages}{97--116} (\bibinfo{year}{2003}).

\bibitem{schools2}
\bibinfo{author}{M.~Klatte}, \bibinfo{author}{J.~Hellbrück},
  \bibinfo{author}{J.~Seidel}, and \bibinfo{author}{P.~Leistner},
  \enquote{\bibinfo{title}{Effects of classroom acoustics on performance and
  well-being in elementary school children: A field study}},
  \bibinfo{journal}{Environ. Behav.} \textbf{42}(5), \bibinfo{pages}{659--692}
  (\bibinfo{year}{2010}).

\bibitem{sundbyberg}
\bibinfo{author}{A.~Seddigh}, \bibinfo{author}{E.~Berntson},
  \bibinfo{author}{F.~Jönsson}, \bibinfo{author}{C.~B. Danielson}, and
  \bibinfo{author}{H.~Westerlund}, \enquote{\bibinfo{title}{The effect of noise
  absorption variation in open-plan offices: A field study with a cross-over
  design}},  \bibinfo{journal}{J. Environ. Psych.} \textbf{44},
  \bibinfo{pages}{34--44} (\bibinfo{year}{2015}).

\bibitem{officestress}
\bibinfo{author}{G.~W. Evans} and \bibinfo{author}{D.~Johnsson},
  \enquote{\bibinfo{title}{Stress and open-office noise}},
  \bibinfo{journal}{J. Appl. Psych.} \textbf{85}(5), \bibinfo{pages}{779--783}
  (\bibinfo{year}{2000}).

\bibitem{FDTD}
\bibinfo{author}{D.~Botteldooren}, \enquote{\bibinfo{title}{Finite-difference
  time-domain simulation of low-frequency room acoustic problems}},
  \bibinfo{journal}{J. Acoust. Soc. Am} \textbf{98}(6),
  \bibinfo{pages}{3302--3308} (\bibinfo{year}{1995}).

\bibitem{FEM}
\bibinfo{author}{A.~Craggs}, \enquote{\bibinfo{title}{A finite element method
  for free vibration of air in ducts and rooms with absorbing walls}},
  \bibinfo{journal}{J. Sound Vib} \textbf{73}(4), \bibinfo{pages}{568--576}
  (\bibinfo{year}{1994}).

\bibitem{SEM}
\bibinfo{author}{F.~Pind}, \bibinfo{author}{A.~P. Engsig-Karup},
  \bibinfo{author}{C.~H. Jeong}, \bibinfo{author}{J.~S. Hesthaven},
  \bibinfo{author}{M.~S. Mejling}, and \bibinfo{author}{J.~S. Andersen},
  \enquote{\bibinfo{title}{Time domain room acoustic simulations using the
  spectral element method}},  \bibinfo{journal}{J. Acoust. Soc. Am}
  \textbf{145}(6), \bibinfo{pages}{3299--3310} (\bibinfo{year}{2019}).

\bibitem{Geometr1}
\bibinfo{author}{L.~Savioja} and \bibinfo{author}{U.~P. Svensson},
  \enquote{\bibinfo{title}{Overview of geometrical room acoustic modeling
  techniques}},  \bibinfo{journal}{JASA} \textbf{138}(2),
  \bibinfo{pages}{708--730} (\bibinfo{year}{2015}).

\bibitem{CRBM}
\bibinfo{author}{J.~S. Hesthaven}, \bibinfo{author}{G.~Rozza}, and
  \bibinfo{author}{B.~Stamm}, \emph{\bibinfo{title}{Certified Reduced Basis
  Methods for Parametrized Partial Differential Equations}}
  (\bibinfo{publisher}{Springer}, \bibinfo{year}{2016}).

\bibitem{parallel}
\bibinfo{author}{A.~Melander}, \bibinfo{author}{E.~S. .~F. Pind},
  \bibinfo{author}{A.~Engsig-Karup}, \bibinfo{author}{C.~Jeong},
  \bibinfo{author}{T.~Warburton}, \bibinfo{author}{N.~Chalmers}, and
  \bibinfo{author}{J.~S. Hesthaven}, \enquote{\bibinfo{title}{Massive parallel
  nodal discontinuous galerkin finite element method simulator for room
  acoustics}},  \bibinfo{journal}{Int. J. of High Performance Computing
  Applications (under review in 2022)}  (\bibinfo{year}{2020}).

\bibitem{Nborrel2}
\bibinfo{author}{N.~Borrel-Jensen}, \bibinfo{author}{A.~Engsig-Karup}, and
  \bibinfo{author}{C.-H. Jeong}, \enquote{\bibinfo{title}{Physics-informed
  neural networks for one-dimensional sound field predictions with
  parameterized sources and impedance boundaries}},  \bibinfo{journal}{JASA
  Express Letters.} \textbf{1}(12), \bibinfo{pages}{122402}
  (\bibinfo{year}{2021}).

\bibitem{MORsota}
\bibinfo{author}{P.~Benner}, \bibinfo{author}{S.~Gugercin}, and
  \bibinfo{author}{K.~Willcox}, \enquote{\bibinfo{title}{A survey of
  projection-based model reduction methods for parametric dynamical systems}},
  \bibinfo{journal}{SIAM} \textbf{57}(4), \bibinfo{pages}{10.1137/130932715}
  (\bibinfo{year}{2015}).

\bibitem{CRBM2}
\bibinfo{author}{A.~Quarteroni}, \bibinfo{author}{G.~Rozza}, and
  \bibinfo{author}{A.~Manzoni}, \enquote{\bibinfo{title}{Certified reduced
  basis approximation for parametrized partial differential equations and
  applications}},  \bibinfo{journal}{J. Math Ind} \textbf{1}(3),
  \bibinfo{pages}{.} (\bibinfo{year}{2011}).

\bibitem{NgocCRBM}
\bibinfo{author}{N.~N. Cuong}, \bibinfo{author}{K.~Veroy}, and
  \bibinfo{author}{A.~T. Patera}, \emph{\bibinfo{title}{Certified real-time
  solution of parametrized partial differential equations}}
  (\bibinfo{publisher}{Handbook of {M}aterials {M}odeling},
  \bibinfo{year}{2005, pp. 1529-1564}).

\bibitem{RozzaHuynhRBM}
\bibinfo{author}{G.~Rozza}, \bibinfo{author}{D.~B.~P. Huynh}, and
  \bibinfo{author}{A.~T. Patera}, \enquote{\bibinfo{title}{Reduced basis
  approximation and a posteriori error estimation for affinely parametrized
  elliptic coercive partial differential equations}},
  \bibinfo{journal}{Archives of Computational Methods in Engineering}
  \textbf{15}(3), \bibinfo{pages}{229--275} (\bibinfo{year}{2008}).

\bibitem{DrohmannRBM}
\bibinfo{author}{M.~Drohmann} and \bibinfo{author}{K.~Carlberg},
  \enquote{\bibinfo{title}{The {R}{O}{M}{E}{S} method for statistical modeling
  of reduced-order-model error}},  \bibinfo{journal}{{S}{I}{A}{M}/{A}{S}{A} J.
  on Uncer. Quant.} \textbf{3}(1), \bibinfo{pages}{116--145}
  (\bibinfo{year}{2015}).

\bibitem{PrudRBM}
\bibinfo{author}{C.~Prud’homme} and \bibinfo{author}{A.~Patera},
  \enquote{\bibinfo{title}{Reduced-basis output bounds for approximately
  parametrized elliptic coercive partial differential equations}},
  \bibinfo{journal}{Computing and Visualization in Science} \textbf{6}(2-3),
  \bibinfo{pages}{147--162} (\bibinfo{year}{2004}).

\bibitem{QuarteroniRBM}
\bibinfo{author}{A.~Quarteroni}, \bibinfo{author}{A.~Manzoni}, and
  \bibinfo{author}{F.~Negri}, \emph{\bibinfo{title}{Reduced {B}asis {M}ethods
  for {P}artial {D}ifferential {E}quations: {A}n {I}ntroduction}}
  (\bibinfo{publisher}{Springer}, \bibinfo{year}{2015}).

\bibitem{HolmesTurb}
\bibinfo{author}{G.~Berkooz}, \bibinfo{author}{P.~Holmes},
  \bibinfo{author}{J.~L. Lumleyand}, and \bibinfo{author}{C.~W. Rowley},
  \emph{\bibinfo{title}{Turbulence, {C}oherent {S}tructures, {D}ynamical
  {S}ystems and {S}ymmetry}}  (\bibinfo{publisher}{Cambridge University Press},
  \bibinfo{year}{1996}).

\bibitem{RBMSuccessLimitations}
\bibinfo{author}{M.~Ohlberger} and \bibinfo{author}{S.~Rave},
  \enxquote{\bibinfo{title}{Reduced basis methods: Success, limitations and
  future challenges}} , \bibinfo{howpublished}{Proceedings Of The Conference
  Algoritmy, 1-12} (\bibinfo{year}{2016}).

\bibitem{Quirin1}
\bibinfo{author}{Q.~Aumanna}, \bibinfo{author}{E.~Deckersb},
  \bibinfo{author}{S.~Jonckheerec}, \bibinfo{author}{W.~Desmetc}, and
  \bibinfo{author}{G.~Muller}, \enquote{\bibinfo{title}{Automatic model order
  reduction for systems with frequency-dependent material properties}},
  \bibinfo{journal}{Comput. Meth. Appl. Mech. Eng.} \textbf{397},
  \bibinfo{pages}{115076} (\bibinfo{year}{2022}).

\bibitem{Quirin2}
\bibinfo{author}{Q.~A.~G. Muller}, \enquote{\bibinfo{title}{Predicting near
  optimal interpolation points for parametric model order reduction using
  regression models}},  \bibinfo{journal}{App. Math and Mech.} \textbf{20}(1),
  \bibinfo{pages}{10} (\bibinfo{year}{2021}).

\bibitem{rombem}
\bibinfo{author}{X.~Xiea}, \bibinfo{author}{W.~Wanga},
  \bibinfo{author}{K.~Hea}, and \bibinfo{author}{G.~Lia},
  \enquote{\bibinfo{title}{Fast model order reduction boundary element method
  for large-scale acoustic systems involving surface impedance}},
  \bibinfo{journal}{Comput. Meth. Appl. Mech. Eng.} \textbf{400},
  \bibinfo{pages}{115618} (\bibinfo{year}{2022}).

\bibitem{Ganeshele}
\bibinfo{author}{M.~Ganesh}, \bibinfo{author}{J.~Hesthaven}, and
  \bibinfo{author}{B.~Stamm}, \enquote{\bibinfo{title}{A reduced basis method
  for electromagnetic scattering by multiple particles in three dimensions}},
  \bibinfo{journal}{J. Comput. Physics} \textbf{231}(23),
  \bibinfo{pages}{7756--7779} (\bibinfo{year}{2012}).

\bibitem{Chenelec}
\bibinfo{author}{Y.~Chen} and \bibinfo{author}{J.~S. Hesthaven},
  \enquote{\bibinfo{title}{Certified reduced basis methods and output bounds
  for the harmonic maxwell’s equations}},  \bibinfo{journal}{SIAM J. Sci.
  Comput.,} \textbf{32}(2), \bibinfo{pages}{970--996} (\bibinfo{year}{2010}).

\bibitem{AmsallemFluid}
\bibinfo{author}{D.~Amsallem}, \bibinfo{author}{J.~Cortial}, and
  \bibinfo{author}{C.~Farhat}, \enquote{\bibinfo{title}{Towards realtime
  computational-fluid-dynamics-based aeroelastic computations using a database
  of reduced-order information}},  \bibinfo{journal}{AIAA J.,} \textbf{48}(9),
  \bibinfo{pages}{2029--2037} (\bibinfo{year}{2010}).

\bibitem{phdGiere}
\bibinfo{author}{S.~Giere}, \enquote{\bibinfo{title}{Numerical and analytical
  aspects of pod-based reduced-order modeling in computational fluid
  dynamics}}, \bibinfo{type}{Ph{D} thesis}, \bibinfo{school}{Free University of
  Berlin, Germany}, \bibinfo{year}{2016}.

\bibitem{phdGrepl}
\bibinfo{author}{M.~A. Grepl}, \enquote{\bibinfo{title}{Reduced-basis
  approximation and a posteriori error estimation for parabolic partial
  differential equations}}, \bibinfo{type}{Ph{D} thesis},
  \bibinfo{school}{Massachusetts Institute of Technology, United States},
  \bibinfo{year}{2005}.

\bibitem{SrinivasaVA}
\bibinfo{author}{R.~S. Puri}, \bibinfo{author}{D.~Morrey},
  \bibinfo{author}{A.~J. Bell}, \bibinfo{author}{J.~F. Durodola},
  \bibinfo{author}{E.~B. Rudnyi}, and \bibinfo{author}{J.~G. Korvink},
  \enquote{\bibinfo{title}{Reduced order fully coupled structural– acoustic
  analysis via implicit moment matching}},  \bibinfo{journal}{Appl. Math. Mod}
  \textbf{33}(11), \bibinfo{pages}{4097--4119} (\bibinfo{year}{2009}).

\bibitem{HetmaniukVA}
\bibinfo{author}{U.~Hetmaniuk}, \bibinfo{author}{R.~Tezaur}, and
  \bibinfo{author}{C.~Farhat}, \enquote{\bibinfo{title}{Review and assessment
  of interpolatory model order reduction methods for frequency response
  structural dynamics and acoustics problems}},  \bibinfo{journal}{Int. J. Num.
  Meth. Eng.} \textbf{90}(13), \bibinfo{pages}{1636--1662}
  (\bibinfo{year}{2012}).

\bibitem{HerrmannVA}
\bibinfo{author}{J.~Herrmann}, \bibinfo{author}{M.~Maess}, and
  \bibinfo{author}{L.~Gaul}, \enquote{\bibinfo{title}{Substructuring including
  interface reduction for the efficient vibro-acoustic simulation of
  fluid-filled piping systems}},  \bibinfo{journal}{Mech. Sys. Sig. Proc}
  \textbf{24}(1), \bibinfo{pages}{153--163} (\bibinfo{year}{2010}).

\bibitem{romart1}
\bibinfo{author}{H.~Basir}, \bibinfo{author}{A.~Javaherian},
  \bibinfo{author}{Z.~Shomali}, \bibinfo{author}{R.~Firouz-Abadi}, and
  \bibinfo{author}{S.~Gholamy}, \enquote{\bibinfo{title}{Acoustic wave
  propagation simulation by reduced order modelling}},
  \bibinfo{journal}{Exploration Geophysics} \textbf{49}(3),
  \bibinfo{pages}{386--397} (\bibinfo{year}{2017}).

\bibitem{romart2}
\bibinfo{author}{L.~Borcea}, \bibinfo{author}{J.~Garnier},
  \bibinfo{author}{A.~Mamonov}, and \bibinfo{author}{J.~Zimmerling},
  \enquote{\bibinfo{title}{Reduced order model approach for imaging with
  waves}},  \bibinfo{journal}{Inverse Problems} \textbf{38}(2),
  \bibinfo{pages}{0025004} (\bibinfo{year}{2022}).

\bibitem{romart3}
\bibinfo{author}{J.~Willhite}, \bibinfo{author}{K.~Frampton}, and
  \bibinfo{author}{D.~W. Grantham}, \enquote{\bibinfo{title}{Reduced order
  modeling of head related transfer functions for virtual acoustic displays}},
  \bibinfo{journal}{JASA} \textbf{113}(4), \bibinfo{pages}{2270--2270}
  (\bibinfo{year}{2003}).

\bibitem{romart4}
\bibinfo{author}{S.~Bukka}, \bibinfo{author}{Y.~Law},
  \bibinfo{author}{H.~Santo}, and \bibinfo{author}{R.~Chan},
  \enquote{\bibinfo{title}{Reduced order model for nonlinear multi-directional
  ocean wave propagation}},  \bibinfo{journal}{Physics of Fluids}
  \textbf{33}(11), \bibinfo{pages}{117115} (\bibinfo{year}{2021}).

\bibitem{romart5}
\bibinfo{author}{A.~Tello} and \bibinfo{author}{R.~Codina},
  \enquote{\bibinfo{title}{Field-to-field coupled fluid structure interaction:
  A reduced order model study}},  \bibinfo{journal}{International Journal for
  Numerical Methods in Engineering} \textbf{122}(1), \bibinfo{pages}{53--81}
  (\bibinfo{year}{2021}).

\bibitem{Miller2}
\bibinfo{author}{M.~Miller}, \bibinfo{author}{S.~Ophem},
  \bibinfo{author}{E.~Deckers}, and \bibinfo{author}{W.~Desmet},
  \enquote{\bibinfo{title}{Time-domain impedance boundary conditions for
  acoustic reduced order finite element simulations}},
  \bibinfo{journal}{Computer Methods in Applied Mechanics and Engineering.}
  \textbf{387}(15), \bibinfo{pages}{114173} (\bibinfo{year}{2021}).

\bibitem{RBMHermes}
\bibinfo{author}{H.~S. Llopis}, \bibinfo{author}{A.~P. Engsig-Karup},
  \bibinfo{author}{C.-H. Jeong}, \bibinfo{author}{F.~Pind}, and
  \bibinfo{author}{J.~S. Hesthaven}, \enquote{\bibinfo{title}{Reduced basis
  methods for numerical room acoustic simulations with parametrized
  boundaries}},  \bibinfo{journal}{JASA} \textbf{152}(2),
  \bibinfo{pages}{851--865} (\bibinfo{year}{2022}).

\bibitem{NODALDGFEM}
\bibinfo{author}{J.~S. Hesthaven} and \bibinfo{author}{T.~Warburton},
  \emph{\bibinfo{title}{Nodal {D}iscontinuous {G}alerkin
  {M}ethods—{A}lgorithms}}  (\bibinfo{publisher}{Springer, New York},
  \bibinfo{year}{2008}).

\bibitem{SPHPAllan}
\bibinfo{author}{H.~Xu}, \bibinfo{author}{C.~Cantwell},
  \bibinfo{author}{C.~Monteserin}, \bibinfo{author}{C.~Eskilsson},
  \bibinfo{author}{A.~P. Engsig-Karup}, and \bibinfo{author}{S.~Sherwin},
  \enquote{\bibinfo{title}{Spectral/hp element methods: Recent developments,
  applications, and perspectives}},  \bibinfo{journal}{Journal of
  Hydrodynamics} \textbf{30}(1), \bibinfo{pages}{1--22} (\bibinfo{year}{2018}).

\bibitem{Cotte}
\bibinfo{author}{B.~Cotté}, \bibinfo{author}{P.~Blanc-Benon},
  \bibinfo{author}{C.~Bogey}, and \bibinfo{author}{F.~Poisson},
  \enquote{\bibinfo{title}{Time-domain impedance boundary conditions for
  simulations of outdoor sound propagation}},  \bibinfo{journal}{AIAA Journal}
  \textbf{47}, \bibinfo{pages}{10} (\bibinfo{year}{2009}).

\bibitem{Troian}
\bibinfo{author}{R.~Troian}, \bibinfo{author}{D.~Dragna},
  \bibinfo{author}{C.~Bailly}, and \bibinfo{author}{M.-A. Galland},
  \enquote{\bibinfo{title}{Broadband liner impedance eduction for multimodal
  acoustic propagation in the presence of a mean flow}},  \bibinfo{journal}{J.
  Sound Vib.} \textbf{392}, \bibinfo{pages}{200--216} (\bibinfo{year}{2017}).

\bibitem{Mikis}
\bibinfo{author}{Y.~Miki}, \enquote{\bibinfo{title}{Acoustical properties of
  porous materials—modifications of delany-bazley models}},
  \bibinfo{journal}{J. Acoust. Soc. Jpn.} \textbf{11}(1),
  \bibinfo{pages}{19.24} (\bibinfo{year}{1990}).

\bibitem{allardbook}
\bibinfo{author}{J.~F. Allard} and \bibinfo{author}{N.~Atalla},
  \emph{\bibinfo{title}{Propagation of Sound in Porous Media: {M}odelling sound
  absorbing materials}}, \bibinfo{}{3rd} ed.  (\bibinfo{publisher}{Wiley, West
  Sussex}, \bibinfo{year}{2009}).

\bibitem{vecfitt}
\bibinfo{author}{B.~Gustavsen} and \bibinfo{author}{A.~Semlyen},
  \enquote{\bibinfo{title}{Rational approximation of frequency domain responses
  by vector fitting}},  \bibinfo{journal}{IEEE Trans. Power Delivery}
  \textbf{14}(3), \bibinfo{pages}{1052–1061} (\bibinfo{year}{1999}).

\bibitem{Bigoni}
\bibinfo{author}{C.~Bigoni} and \bibinfo{author}{J.~S. Hesthaven},
  \enquote{\bibinfo{title}{Simulation-based anomaly detection and damage
  localization: an application to structural health monitoring}},
  \bibinfo{journal}{Comput. Meth. Appl. Mech. Eng} \textbf{63},
  \bibinfo{pages}{12896} (\bibinfo{year}{2020}).

\bibitem{Weeks}
\bibinfo{author}{W.~T. Weeks}, \enquote{\bibinfo{title}{Numerical inversion of
  laplace transform using {L}aguerre functions}},  \bibinfo{journal}{J. Assc.
  Comp. Mach.} \textbf{13}(3), \bibinfo{pages}{419--429}
  (\bibinfo{year}{1966}).

\bibitem{symplectic}
\bibinfo{author}{L.~Peng} and \bibinfo{author}{K.~Mohseni},
  \enquote{\bibinfo{title}{Symplectic model reduction of hamiltonian systems}},
   \bibinfo{journal}{SIAM J. Sci. Comput.} \textbf{38}(1),
  \bibinfo{pages}{A1--A27} (\bibinfo{year}{2016}).

\bibitem{Sakamoto}
\bibinfo{author}{S.~Sakamoto}, \enquote{\bibinfo{title}{Phase-error analysis of
  high-order finite difference time domain scheme and its influence on
  calculation results of impulse response in closed sound field}},
  \bibinfo{journal}{Acoust. Sci.Tech} \textbf{28}(5),
  \bibinfo{pages}{295–309} (\bibinfo{year}{2007}).

\bibitem{GRCH}
\bibinfo{author}{C.~Jeong} and \bibinfo{author}{J.~lh},
  \enquote{\bibinfo{title}{Effects of source and receiver locations in
  predicting room transfer functions by a phased beam tracing method}},
  \bibinfo{journal}{J. Acoust. Soc. Am} \textbf{131}(5),
  \bibinfo{pages}{3864--3875} (\bibinfo{year}{2012}).

\bibitem{cox}
\bibinfo{author}{T.~Cox} and \bibinfo{author}{P.~DAntonio},
  \emph{\bibinfo{title}{Acoustic absorbers and diffusers : theory, design and
  application : Theory, design and application}}  (\bibinfo{publisher}{Spon},
  \bibinfo{year}{2004}).

\bibitem{ISO3382-1}
\enquote{\bibinfo{title}{{EN ISO 3382-1:2009, Measurement of room acoustic
  parameters, part 1: Performance spaces.}}}, \bibinfo{type}{Standard},
  \bibinfo{institution}{International Organization for Standardization}
  (\bibinfo{year}{2009}).

\bibitem{jnd6}
\bibinfo{author}{P.~Luizard}, \bibinfo{author}{B.~F.~G. Katz}, and
  \bibinfo{author}{C.~Guastavino}, \enquote{\bibinfo{title}{Perceptual
  thresholds for realistic double-slope decay reverberation in large coupled
  spaces}},  \bibinfo{journal}{J. Acoust. Soc. Am} \textbf{137}(1),
  \bibinfo{pages}{75–84} (\bibinfo{year}{2015}).

\bibitem{jnd2}
\bibinfo{author}{M.~G. Blevins}, \bibinfo{author}{A.~T. Buck},
  \bibinfo{author}{Z.~Peng}, and \bibinfo{author}{L.~M. Wang},
  \enxquote{\bibinfo{title}{Quantifying the just noticeable difference of
  reverberation time with band-limited noise centered around 1000 hz using a
  transformed up-down adaptive method}} , \bibinfo{howpublished}{in Proc. Int.
  Symp. Room Acoustics (ISRA) Toronto, Canada} (\bibinfo{year}{2013}).

\bibitem{jnd4}
\bibinfo{author}{M.~Karjalainen} and \bibinfo{author}{H.~J{\"a}rvel{\"a}inen},
  \enxquote{\bibinfo{title}{More about this reverberation science: Perceptually
  good late reverberation}} , \bibinfo{howpublished}{In Proc. 111th Audio Eng.
  Soc. Conv., New York, USA} (\bibinfo{year}{2001}).

\bibitem{jnd5}
\bibinfo{author}{Z.~Meng}, \bibinfo{author}{F.~Zhao}, and
  \bibinfo{author}{M.~He}, \enxquote{\bibinfo{title}{The just noticeable
  difference of noise length and reverberation perception}} ,
  \bibinfo{howpublished}{In Proc. Int. Symp. Com. Inf. Tech., Bangkok,
  Thailand} (\bibinfo{year}{2006}).

\bibitem{jnd3}
\bibinfo{author}{K.~Prawda}, \bibinfo{author}{S.~J. Schlecht}, and
  \bibinfo{author}{V.~V{\"a}lim{\"a}ki}, \enxquote{\bibinfo{title}{Improved
  reverberation time control for feedback delay networks}} ,
  \bibinfo{howpublished}{In Proc. Int. Conf. Digital Audio Effects, Birmingham,
  UK} (\bibinfo{year}{2019}).

\end{thebibliography}


\appendix
\section*{Appendix}
\setcounter{table}{0}
\renewcommand{\thetable}{A\arabic{table}}
\begin{table}[h]
 \centering
\caption{Parameters selected to build the 2D ROM. Each row corresponds to a FOM simulation with the corresponding parameters marked for each column whose values are presented in Table \ref{Tab:ROM2d}.}
\label{Tab:ROM2dparam}
\resizebox{.51\textwidth}{!}{
\begin{tabular}{|l|l|l|l|l|l|l|l|l|l|l|l|l|l|l|l|l|l|l|l|l|l|}
\hline
\multicolumn{1}{|c|}{} & \multicolumn{1}{c|}{CE$_{\sigma1}$} & \multicolumn{1}{c|}{CE$_{\sigma2}$}& CE$_{\sigma3}$ & CE$_{d1}$ & CE$_{d2}$ & CE$_{d3}$ & CE$_{d_01}$ & CE$_{d_02}$ & CE$_{d_03}$ & FL$_{Z1}$ & FL$_{Z2}$ & FL$_{Z3}$ & FL$_{\sigma1}$ & FL$_{\sigma2}$ & FL$_{\sigma3}$ & WL$_{\sigma1}$ & WL$_{\sigma2}$ & WL$_{\sigma3}$ & WR$_{\sigma1}$ & WR$_{\sigma2}$ & WR$_{\sigma3}$ \\ \hline \hline
FOM$_1$  & \multicolumn{1}{c|}{X} &  &  &  & \multicolumn{1}{c|}{X} &  &  & \multicolumn{1}{c|}{X} &  &  & &  &  & \multicolumn{1}{c|}{X} &  &  &\multicolumn{1}{c|}{X}  &  &  & \multicolumn{1}{c|}{X} &  \\ \hline\hline
FOM$_2$  &  &\multicolumn{1}{c|}{X}  &  &  & \multicolumn{1}{c|}{X} &  &  & \multicolumn{1}{c|}{X} &  &  & &  &  & \multicolumn{1}{c|}{X} &  &  &\multicolumn{1}{c|}{X}  &  &  & \multicolumn{1}{c|}{X} &  \\ \hline \hline
FOM$_3$  &  & & \multicolumn{1}{c|}{X}  & & \multicolumn{1}{c|}{X} &  &  & \multicolumn{1}{c|}{X} &  &  & &  &  & \multicolumn{1}{c|}{X} &  &  &\multicolumn{1}{c|}{X}  &  &  & \multicolumn{1}{c|}{X} &  \\ \hline \hline
FOM$_4$  &  & \multicolumn{1}{c|}{X}  &  & \multicolumn{1}{c|}{X}  &  &  &  & \multicolumn{1}{c|}{X} &  &  & &  &  & \multicolumn{1}{c|}{X} &  &  &\multicolumn{1}{c|}{X}  &  &  & \multicolumn{1}{c|}{X} &  \\ \hline \hline
FOM$_5$  &  & \multicolumn{1}{c|}{X}  &  &  & \multicolumn{1}{c|}{X}   &  &  & \multicolumn{1}{c|}{X} &  &  & &  &  & \multicolumn{1}{c|}{X} &  &  &\multicolumn{1}{c|}{X}  &  &  & \multicolumn{1}{c|}{X} &  \\ \hline \hline
FOM$_6$  &  & \multicolumn{1}{c|}{X}  &  &  &   & \multicolumn{1}{c|}{X}  &  & \multicolumn{1}{c|}{X} &  &  & &  &  & \multicolumn{1}{c|}{X} &  &  &\multicolumn{1}{c|}{X}  &  &  & \multicolumn{1}{c|}{X} &  \\ \hline \hline
FOM$_7$  &  & \multicolumn{1}{c|}{X}  &  &  &\multicolumn{1}{c|}{X}   &   & \multicolumn{1}{c|}{X} & &  &  & &  &  & \multicolumn{1}{c|}{X} &  &  &\multicolumn{1}{c|}{X}  &  &  & \multicolumn{1}{c|}{X} &  \\ \hline \hline
FOM$_8$  &  & \multicolumn{1}{c|}{X}  &  &  &\multicolumn{1}{c|}{X}   &   &  &\multicolumn{1}{c|}{X} &  &  & &  &  & \multicolumn{1}{c|}{X} &  &  &\multicolumn{1}{c|}{X}  &  &  & \multicolumn{1}{c|}{X} &  \\ \hline \hline
FOM$_9$  &  & \multicolumn{1}{c|}{X}  &  &  &\multicolumn{1}{c|}{X}   &   &  & &\multicolumn{1}{c|}{X}  &  & &  &  & \multicolumn{1}{c|}{X} &  &  &\multicolumn{1}{c|}{X}  &  &  & \multicolumn{1}{c|}{X} &  \\ \hline \hline
FOM$_{10}$ &  & \multicolumn{1}{c|}{X}  &  &  &\multicolumn{1}{c|}{X}   &   &  &\multicolumn{1}{c|}{X}  & & \multicolumn{1}{c|}{X} & &  &  & &  &  &\multicolumn{1}{c|}{X}  &  &  & \multicolumn{1}{c|}{X} &  \\ \hline \hline
FOM$_{11}$ &  & \multicolumn{1}{c|}{X}  &  &  &\multicolumn{1}{c|}{X}   &   &  &\multicolumn{1}{c|}{X}  & & & \multicolumn{1}{c|}{X} &  &  &  &  &  &\multicolumn{1}{c|}{X}  &  &  & \multicolumn{1}{c|}{X} &  \\ \hline \hline
FOM$_{12}$ &  & \multicolumn{1}{c|}{X}  &  &  &\multicolumn{1}{c|}{X}   &   &  &\multicolumn{1}{c|}{X}  & & &  & \multicolumn{1}{c|}{X} &  &  &  &  &\multicolumn{1}{c|}{X}  &  &  & \multicolumn{1}{c|}{X} &  \\ \hline \hline
FOM$_{13}$  &  & \multicolumn{1}{c|}{X}  &  &  &\multicolumn{1}{c|}{X}   &   &  &\multicolumn{1}{c|}{X}  & & &  &  &\multicolumn{1}{c|}{X}  &  &  &  &\multicolumn{1}{c|}{X}  &  &  & \multicolumn{1}{c|}{X} &  \\ \hline \hline
FOM$_{14}$ &  & \multicolumn{1}{c|}{X}  &  &  &\multicolumn{1}{c|}{X}   &   &  &\multicolumn{1}{c|}{X}  & & &  &  & &\multicolumn{1}{c|}{X}   &  &  &\multicolumn{1}{c|}{X}  &  &  & \multicolumn{1}{c|}{X} &  \\ \hline \hline
FOM$_{15}$&  & \multicolumn{1}{c|}{X}  &  &  &\multicolumn{1}{c|}{X}   &   &  &\multicolumn{1}{c|}{X}  & & &  &  & &  &\multicolumn{1}{c|}{X}   &  &\multicolumn{1}{c|}{X}  &  &  & \multicolumn{1}{c|}{X} &  \\ \hline \hline
FOM$_{16}$ &  & \multicolumn{1}{c|}{X}  &  &  &\multicolumn{1}{c|}{X}   &   &  &\multicolumn{1}{c|}{X}  & & &  &  & & \multicolumn{1}{c|}{X} &   &\multicolumn{1}{c|}{X}  &  &  &  & \multicolumn{1}{c|}{X} &  \\ \hline \hline
FOM$_{17}$ &  & \multicolumn{1}{c|}{X}  &  &  &\multicolumn{1}{c|}{X}   &   &  &\multicolumn{1}{c|}{X}  & & &  &  & & \multicolumn{1}{c|}{X} &   & & \multicolumn{1}{c|}{X}  &  &  & \multicolumn{1}{c|}{X} &  \\ \hline \hline
FOM$_{18}$ &  & \multicolumn{1}{c|}{X}  &  &  &\multicolumn{1}{c|}{X}   &   &  &\multicolumn{1}{c|}{X}  & & &  &  &  &\multicolumn{1}{c|}{X}   & &   &\multicolumn{1}{c|}{X}  &  \multicolumn{1}{c|}{X}&  & & \\ \hline \hline
FOM$_{19}$ &  & \multicolumn{1}{c|}{X}  &  &  &\multicolumn{1}{c|}{X}   &   &  &\multicolumn{1}{c|}{X}  & & &  &  & &\multicolumn{1}{c|}{X}   & &  &\multicolumn{1}{c|}{X}  & & \multicolumn{1}{c|}{X}  & & \\ \hline \hline
FOM$_{20}$  &  & \multicolumn{1}{c|}{X}  &  &  &\multicolumn{1}{c|}{X}   &   &  &\multicolumn{1}{c|}{X}  & & &  &  &&\multicolumn{1}{c|}{X}   & & & \multicolumn{1}{c|}{X} & &  &\multicolumn{1}{c|}{X}  & \\ \hline \hline
FOM$_{21}$ &  & \multicolumn{1}{c|}{X}  &  &  &\multicolumn{1}{c|}{X}   &   &  &\multicolumn{1}{c|}{X}  & & &  &  &  &\multicolumn{1}{c|}{X}   & &  &\multicolumn{1}{c|}{X}  & &  &  &\multicolumn{1}{c|}{X} \\ \hline \hline
\end{tabular}
}
\end{table}

\begin{table}[h]
 \centering
\caption{Parameters selected to build the 3D ROM. Each row corresponds to a FOM simulation with the corresponding parameters marked for each column whose values are presented in Table \ref{Tab:ROM3d}.}
\label{Tab:ROM3dparam}
\resizebox{\columnwidth}{!}{
\begin{tabular}{|l|l|l|l|l|l|l|l|l|l|l|l|l|}
\hline
\multicolumn{1}{|c|}{} & \multicolumn{1}{c|}{CE$_{\sigma1}$} & \multicolumn{1}{c|}{CE$_{\sigma2}$}& CE$_{\sigma3}$ & CE$_{d_01}$ & CE$_{d_02}$ & CE$_{d_03}$ & WW$_{\sigma1}$ &  WW$_{\sigma2}$ & WW$_{\sigma3}$ & WN$_{d_01}$ & WN$_{d_02}$ & WN$_{d_03}$  \\ \hline \hline
\multicolumn{1}{|c}{FOM$_1$} &\multicolumn{1}{|c|}{X} & & & &\multicolumn{1}{c|}{X} & & &\multicolumn{1}{c|}{X} & & &\multicolumn{1}{c|}{X} & \\ \hline \hline
\multicolumn{1}{|c|}{FOM$_2$}& & \multicolumn{1}{c|}{X}& & &\multicolumn{1}{c|}{X} & & &\multicolumn{1}{c|}{X} & & &\multicolumn{1}{c|}{X} & \\ \hline \hline
\multicolumn{1}{|c|}{FOM$_3$} & & &\multicolumn{1}{c|}{X}  & &\multicolumn{1}{c|}{X} & & &\multicolumn{1}{c|}{X} & & &\multicolumn{1}{c|}{X} & \\ \hline \hline
\multicolumn{1}{|c|}{FOM$_4$} & &\multicolumn{1}{c|}{X} &  & \multicolumn{1}{c|}{X}& & & &\multicolumn{1}{c|}{X} & & &\multicolumn{1}{c|}{X} & \\ \hline \hline
\multicolumn{1}{|c|}{FOM$_5$} & &\multicolumn{1}{c|}{X} &  & &\multicolumn{1}{c|}{X} & & &\multicolumn{1}{c|}{X} & & &\multicolumn{1}{c|}{X} & \\ \hline \hline
\multicolumn{1}{|c|}{FOM$_6$} & &\multicolumn{1}{c|}{X} &  & & &\multicolumn{1}{c|}{X} & &\multicolumn{1}{c|}{X} & & &\multicolumn{1}{c|}{X} & \\ \hline \hline
\multicolumn{1}{|c|}{FOM$_7$} & &\multicolumn{1}{c|}{X} &  & &\multicolumn{1}{c|}{X} & &\multicolumn{1}{c|}{X} & & & &\multicolumn{1}{c|}{X} & \\ \hline \hline
\multicolumn{1}{|c|}{FOM$_8$}  & &\multicolumn{1}{c|}{X} &  & &\multicolumn{1}{c|}{X} & & &\multicolumn{1}{c|}{X} & & &\multicolumn{1}{c|}{X} & \\ \hline \hline
\multicolumn{1}{|c|}{FOM$_9$} & &\multicolumn{1}{c|}{X} &  & &\multicolumn{1}{c|}{X} & & & &\multicolumn{1}{c|}{X} & &\multicolumn{1}{c|}{X} & \\ \hline \hline
\multicolumn{1}{|c|}{FOM$_{10}$}& &\multicolumn{1}{c|}{X} &  & &\multicolumn{1}{c|}{X} & & &\multicolumn{1}{c|}{X} & & \multicolumn{1}{c|}{X}& & \\ \hline \hline
\multicolumn{1}{|c|}{FOM$_{11}$} & &\multicolumn{1}{c|}{X} &  & &\multicolumn{1}{c|}{X} & & &\multicolumn{1}{c|}{X} & & &\multicolumn{1}{c|}{X} & \\ \hline \hline
\multicolumn{1}{|c|}{FOM$_{12}$}  & &\multicolumn{1}{c|}{X} &  & &\multicolumn{1}{c|}{X} & & &\multicolumn{1}{c|}{X} & & & &\multicolumn{1}{c|}{X} \\ \hline \hline

\end{tabular}
}
\end{table}


\end{document}